\documentstyle[twocolumn,prb,aps,epsfig,amssym]{revtex}

\begin{document}
\draft
\preprint{}

\newcommand{\1}{{\bf \scriptstyle 1}\!\!{1}}
\newcommand{\p}{\partial}
\newcommand{\D}{^{\dagger}}
\newcommand{\bx}{{\bf x}}
\newcommand{\bk}{{\bf k}}
\newcommand{\bv}{{\bf v}}
\newcommand{\bp}{{\bf p}}
\newcommand{\bu}{{\bf u}}
\newcommand{\bA}{{\bf A}}
\newcommand{\bB}{{\bf B}}
\newcommand{\bK}{{\bf K}}
\newcommand{\bL}{{\bf L}}
\newcommand{\bP}{{\bf P}}
\newcommand{\bQ}{{\bf Q}}
\newcommand{\bS}{{\bf S}}
\newcommand{\balpha}{\mbox{\boldmath $\alpha$}}
\newcommand{\bsigma}{\mbox{\boldmath $\sigma$}}
\newcommand{\bSigma}{\mbox{\boldmath $\Sigma$}}
\newcommand{\bomega}{\mbox{\boldmath $\omega$}}
\newcommand{\bpi}{\mbox{\boldmath $\pi$}}
\newcommand{\bphi}{\mbox{\boldmath $\phi$}}
\newcommand{\bnabla}{\mbox{\boldmath $\nabla$}}
\newcommand{\bmu}{\mbox{\boldmath $\mu$}}
\newcommand{\bepsilon}{\mbox{\boldmath $\epsilon$}}

\newcommand{\iLambda}{{\it \Lambda}}
\newcommand{\cL}{{\cal L}}
\newcommand{\cH}{{\cal H}}
\newcommand{\cU}{{\cal U}}

\newcommand{\be}{\begin{equation}}
\newcommand{\ee}{\end{equation}}
\newcommand{\bea}{\begin{eqnarray}}
\newcommand{\eea}{\end{eqnarray}}
\newcommand{\beqa}{\begin{eqnarray*}}
\newcommand{\eeqa}{\end{eqnarray*}}
\newcommand{\nn}{\nonumber}
\newcommand{\DD}{\displaystyle}

\newcommand{\ba}{\left[\begin{array}{c}}
\newcommand{\baa}{\left[\begin{array}{cc}}
\newcommand{\baaa}{\left[\begin{array}{ccc}}
\newcommand{\baaaa}{\left[\begin{array}{cccc}}
\newcommand{\ea}{\end{array}\right]}

\twocolumn[
\hsize\textwidth\columnwidth\hsize\csname
@twocolumnfalse\endcsname

\title{Spin tunneling and phonon-assisted relaxation in Mn$_{12}$-acetate}

\author{Michael N.~Leuenberger\cite{email1} and Daniel Loss\cite{email2}}
\address{Department of Physics and Astronomy, University of Basel \\
Klingelbergstrasse 82, 4056 Basel, Switzerland}

\maketitle

\begin{abstract}
We present a comprehensive theory of the magnetization relaxation in a Mn$_{12}$-acetate crystal in the high-temperature regime ($T\gtrsim 1$ K), which is based on phonon-assisted spin tunneling induced by quartic magnetic anisotropy and weak transverse magnetic fields. The overall relaxation rate as function of the longitudinal magnetic field is calculated and shown to agree well with experimental data including all resonance peaks measured so far. The Lorentzian shape of the resonances, which we obtain via a generalized master equation that includes spin tunneling, is also in good agreement with recent data. We derive a general formula for the tunnel splitting energy of these resonances. We show that fourth-order diagonal terms in the Hamiltonian lead to satellite peaks. A derivation of the effective linewidth of a resonance peak is given and shown to agree well with experimental data. In addition, previously unknown spin-phonon coupling constants are calculated explicitly. The values obtained for these constants and for the sound velocity are also in good agreement with recent data. We show that the spin relaxation in Mn$_{12}$-acetate takes place via several transition paths of comparable weight. These transition paths are expressed in terms of intermediate relaxation times, which are calculated and which can be tested experimentally.

\end{abstract}

]
\narrowtext

\section{Introduction}
\label{intro}

The magnetization relaxation in the molecular magnet Mn$_{12}$-acetate with chemical formula, $\rm [Mn_{12}(CH_3COO)_{16}$ $\rm (H_2O)_4O_{12}]\cdot$ $\rm 2CH_3COOH\cdot 4H_2O$, (henceforth abbreviated as Mn$_{12}$) has attracted much recent interest since several experiments \cite{Paulsen,Paulsen et al,Novak,Sessoli,Novak et al} have indicated unusually long relaxation times --- about two months at a temperature of about 2 K --- as well as pronounced peaks in the relaxation time\cite{Friedman,Thomas,Hernandez} in response to a varying magnetic field $H_z$ when applied along the easy axis of the  Mn$_{12}$ crystal. These peaks correspond to an increased relaxation rate of the magnetization of Mn$_{12}$ and occur when $H_z$ is tuned to multiples of about 0.44 T. According to earlier suggestions\cite{Barbara,Novak2} this phenomenon has been interpreted as a manifestation of resonant tunneling of the magnetization, often referred to as macroscopic quantum tunneling (MQT). 
A qualitative explanation goes as follows. From the microscopic point of view a Mn$_{12}$ cluster acts like a giant spin with length $s=10$ as long as the external magnetic field is small compared to the exchange interactions between the Mn ions, which is fulfilled in the experimental range considered in this paper.
The relaxation rate of the magnetization increases at field values where the spin states become pairwise degenerate. It is this degeneracy that determines the resonance condition.   
As the external field $H_z$ is moved away from a resonance the spin states are no longer perfectly degenerate, and therefore the tunneling probability becomes smaller and thus the relaxation rate. Since the spin system couples to the environmental phonons of the Mn$_{12}$ crystal, the energy levels of the spin states are smeared out. This leads to homogeneously broadened resonance peaks that are of Lorentzian shape. There are also other sources which lead to broadening of the resonances, such as hyperfine and dipolar fields.\cite{FRIEDMAN} They give rise to inhomogeneous broadening with Gaussian-shaped peaks.\cite{Luis,Fernandez} However, this stands in contrast to the measured resonance peaks, which are nearly perfect Lorentzians.\cite{FRIEDMAN} Furthermore, the width of the hyperfine induced Gaussians\cite{Luis,VILLAIN} turns out to be smaller for $T\gtrsim 1$ K than the width of the Lorentzians obtained below and seen in the experiment.\cite{Wernsdorfer} Similarly, dipolar interactions have been ruled out by experiments on diluted samples.\cite{SESSOLI} Thus, for temperatures $T\gtrsim 1$ K we can safely neglect hyperfine and dipolar fields, and the dominant source of the peak broadening can be explained consistently by spin-phonon effects only. 

In a critical comparison 
between model calcu\-lations\cite{Luis,VILLAIN,Villain,Fort,GARANIN,Gunther} and experimental data\cite{Friedman,Thomas,FRIEDMAN} Friedman {\it et al.}\cite{FRIEDMAN} point out that a consistent explanation of the experimentally observed relaxation is still missing. A good starting point for theoretical calculations has been formulated  
by Villain {\it et al.},\cite{Villain} where the relaxation is described in terms of spin-phonon interaction and a generalized Orbach process. However, this approach does not include the dependence on the external field $H_z$. Also, one of the main challenges for theory is to explain the overall shape of the relaxation curve as well as the nearly perfect Lorentzian shape of the measured resonance peaks.\cite{FRIEDMAN}

In this work we perform a model calculation of the magnetization relaxation which is based on phonon-assisted tunneling. We present a self-consistent theory which is for the first time in reasonably good agreement both with the overall relaxation rate (including all resonances) measured by Thomas {\it et al.}\cite{Thomas} (see Fig.~\ref{overall}) and with the Lorentzian shape of the first resonance peaks (see Figs.~\ref{singlepeak} and \ref{fourpeaks}) measured by Friedman {\it et al.}\cite{FRIEDMAN} with high precision for four different temperatures.

Our model, which is introduced in Sec.~\ref{model}, contains five independent parameters: three anisotropy constants $A\gg B\gg B_4$, the misalignment angle $\theta$ (angle between field direction and easy axis, the latter being taken along the $z$ axis), and the sound velocity $c$. The anisotropy constant $B_4$ and the angle $\theta$ are responsible for the spin tunneling. This will be explained in Sec.~\ref{master}. Moreover, we derive the spin-phonon coupling constants in Sec.~\ref{model}. It turns out from our calculations that these constants can be expressed in terms of the anisotropy $A$. The constants $A,B,B_4$ have already been measured\cite{Barra,Zhong} and are known within some experimental uncertainty. We achieve optimal agreement between our theory and data if we proceed as follows. In accordance with Ref.~\onlinecite{FRIEDMAN} we set $\theta = 1^\circ$, while the constants $A,B,B_4$ are fitted to the relaxation data by observing, however, the constraints that $A,B,B_4$ are allowed to vary only within the range of their  experimental uncertainties. The sound velocity $c$ has not been directly measured yet (to our knowledge). However, specific heat measurements\cite{Gomes} yield the Debye temperature of Mn$_{12}$, from which a sound velocity can be deduced that is in excellent agreement with our fit of the sound velocity $c=(1.45-2.0)\times 10^3$ m/s (see Sec.~\ref{relax}). Thus, in contrast to previous results\cite{Luis,VILLAIN,Villain,Fort,GARANIN} our theory is in reasonably good agreement not only with the relaxation data\cite{Thomas,FRIEDMAN} but also with {\it all} experimental parameter values known so far (see Figs.~\ref{overall}, \ref{singlepeak}, and \ref{fourpeaks}). In addition, new predictions are made which can be tested experimentally: the sound velocity $c$ and the intermediate relaxation times $\tau_n$, as well as satellite peaks.

In Sec.~\ref{master}, extending previous work,\cite{Luis,VILLAIN,Villain,Fort,GARANIN} we make use of a generalized master equation which treats phonon-induced spin transitions between nearest and next-nearest energy levels as well as  resonant tunneling due to quartic anisotropies and transverse fields on the same footing, which results in the Lorentzian shape of the resonances.  
We derive the effective linewidth of the Lorentzian peaks (see Sec.~\ref{lorentzian}) as well as a generalized formula of the tunnel splitting energy (see Sec.~\ref{master}). In Sec.~\ref{relax}, we obtain the relaxation time by exactly diagonalizing the master equation. In Sec.~\ref{paths}, solving the master equation analytically, we identify the dominant transition paths (see Figs.~\ref{diagram1} and \ref{diag1ser}) and show that the magnetization reversal is not dominated by just one single path but rather by several paths which can be of comparable weight. We finally note that some of the results of the present paper have been published in Ref.~\onlinecite{LeuenbergerLoss} in a short and less general form. Here we present details of the derivation of these results and generalize them in various ways, leading to new results such as satellite peaks in the overall relaxation curve, relaxation time of an individual relaxation path, an analytical expression for the effective linewidths, and a generalized tunnel splitting formula.


\section{Model}
\label{model}

In accordance with earlier work\cite{Luis,VILLAIN,Villain,Fort,GARANIN,HernandezHxTunneling} we use a single-spin Hamiltonian $\cH=\cH_{\rm a}+\cH_{\rm Z}+\cH_{\rm sp}+\cH_{\rm T}$ including spin-phonon coupling. This model turns out to be sufficient to describe the behavior of the Mn$_{12}$-acetate molecule (for temperatures $T\gtrsim 1$ K). In particular, 
\be 
\cH_{\rm a}=-AS_z^2-BS_z^4 
\label{H_a} 
\ee 
represents the magnetic anisotropy where $A\gg B> 0$. The anisotropy $-AS_z^2$ is  depicted in Fig.~\ref{anisotropy-fig}. We define the easy axis to lie along the $z$ direction.

\begin{figure}[htb]
  \begin{center}
    \leavevmode
\epsfxsize=8.5cm
\epsffile{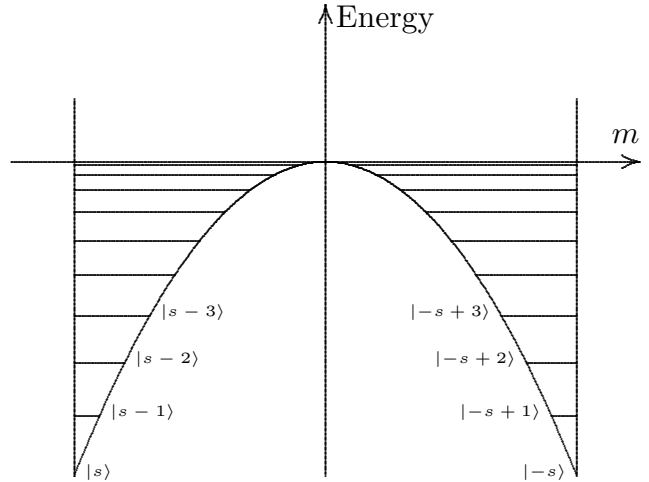}
  \end{center}
\caption{Anisotropy energy $-Am^2-Bm^4$.}
\label{anisotropy-fig}
\end{figure}

Here, ${\bf S}$ is the spin operator with $s=10$, and $A/k_B=0.52-0.56$ K, \cite{Barra,Zhong} and $B/k_B=(1.1-1.3)\times 10^{-3}$ K (Refs.~\cite{Barra} and \cite{Zhong}) are the anisotropy constants ($k_B$ is the Boltzmann factor). The Zeeman term
\be
\cH_{\rm Z}=g\mu_BH_zS_z
\ee
describes the coupling between the external magnetic field $H_z$ and the  spin ${\bf S}$. The $g$ factor is known to be $g=1.9$.\cite{gSessoli}

We denote by $\left|m\right>$, $-s\leq m\leq s$, the eigenstates of ${\cal H}_{\rm a}+{\cal H}_{\rm Z}$ with eigenvalue 
\be
\varepsilon_{m}=-Am^2-Bm^4+g\mu_BH_zm.
\ee 
If the external magnetic field $H_z$ is increased, one obtains doubly degenerate spin states whenever a level $m$ coincides with a level $m'$ on the opposite side of the well (separated by the barrier given by $A$). The resonance condition for double degeneracy, i.e., $\varepsilon_{m}=\varepsilon_{m'}$,  leads to the resonance field
\be
H_z^{mm'}=\frac{n}{g\mu_B}\left[A+B\left(m^2+m'^2\right)\right].
\label{condition}
\ee
As usual, we refer to $n=m+m'=$ even (odd) as even (odd) resonances.

The Hamiltonian 
\be
{\cal H}_{\rm T}=-\frac{1}{2}B_4\left(S_+^4+S_-^4\right)+g\mu_BH_xS_x\, ,
\label{H_T}
\ee
makes tunneling between $S_z$ states possible, where $S_\pm=S_x\pm i S_y$, and $B_4$ is the fourth-order anisotropy constant. $H_x=|{\bf H}|\sin\theta$ is the transverse field, with $\theta$ being the misalignment angle. $H_x$ is assumed to be much smaller than $H_z$, i.e., $\theta$ is at most a few degrees. From experiments\cite{Barra} it is known that $B_4/k_B=(4.3-14.4)\times 10^{-5}$ K. Finally, the most general spin-phonon coupling\cite{Callen} which is allowed by the tetragonal symmetry of the Mn$_{12}$ crystal in leading order is given by
\bea
{\cal H}_{\rm
sp}& = &
g_1(\epsilon_{xx}-\epsilon_{yy})\otimes
(S_x^2-S_y^2)+
\frac{1}{2}g_2\epsilon_{x
y}\otimes\{S_x,S_y\} \nonumber \\
& &
+\left.\frac{1}{2}g_3(\epsilon_{xz}\otimes
\{S_x,S_z\}+\epsilon_{yz}
\otimes\{S_y,S_z\}) \right.\nonumber\\
& &
+\left.\frac{1}{2}g_4(\omega_{xz}\otimes
\{S_x,S_z\}+\omega_{yz}\otimes
\{S_y,S_
z\})\, , \right.
\label{H_sp} \\
& = &
\frac{1}{2}g_1(\epsilon_{xx}-\epsilon_{yy})\otimes
(S_+^2+S_-^2)+
\frac{ i}{4}g_2\epsilon_{xy}\otimes (S_-^2-S_+^2) \nn\\
& &
+\left.\frac{1}{4}g_3[(\epsilon_{xz}- i\epsilon_{yz})
\otimes\{S_+,S_z\}\right.\nn\\ & &
+\left.(\epsilon_{xz}+ i\epsilon_{yz})\otimes\{S_-,S_z\}]
\right.\nn\\ &
&
+\left.\frac{1}{4}g_4[(\omega_{xz}- i\omega_{yz})\otimes\{S_+,S_z\}\right.\nn\\
& & +\left. (\omega_{xz}+ i\omega_{yz})\otimes\{S_-,S_z\}]
\right.,
\label{Spin-phonon}
\eea
where $g_i$, $i=1,2,3,4$, are the spin-phonon coupling constants, which we shall determine in the following.

The linear strain tensor is defined by $\epsilon={\bf \nabla u}$, where  ${\bf u}(x,y,z)$ is the displacement field. Symmetrization of the strain tensor yields
\be
\epsilon_{\alpha\beta}=\frac{1}{2}\left(\frac{\partial
u_\alpha}{\partial\beta}
	+\frac{\partial
u_\beta}{\partial\alpha}\right),
\ee
while the antisymmetrized linear strain tensor reads
\be
\omega_{\alpha\beta}=\frac{1}{2}\left(\frac{\partial
u_\alpha}{\partial\beta}
	-\frac{\partial
u_\beta}{\partial\alpha}\right),
\ee
with $\alpha,\beta=x,y,z$. To determine  $g_i$ occurring in Eq.~(\ref{H_sp}) we follow Dohm and Fulde.\cite{Dohm}
The displacement
\be
{\bf u}=\delta\mbox{\boldmath $\phi$}\times{\bf x}
\label{rot}
\ee
(in leading order) is generated by rotation only. The infinitesimal rotation angle can be calculated by acting with ${\bnabla}_\bx$ (with respect to the position ${\bf x}$) on both sides of Eq.~(\ref{rot}),
\be
\delta\bphi=\frac{1}{2}{\bnabla}\times\bu=\ba \omega_{yz}
\\ \omega_{zx} \\
\omega_{xy}\ea.
\ee
Applying infinitesimal rotations on the spin vector ${\bf S}$
\bea
\lefteqn{
\baaa 1 & 0 & 0 \\ 0 & 1 &
\omega_{yz} \\ 0 & -\omega_{yz} & 1 \ea
\baaa 1 & 0 & \omega_{xz} \\ 0 & 1
& 0 \\ -\omega_{xz} & 0 & 1 \ea
\ba S_x \\ S_y \\ S_z \ea}
\hspace{2.5cm}\nn\\
& & =\ba S_x+\omega_{xz}S_z \\
S_y-\omega_{xz}\omega_{yz}S_x+\omega_{yz}S_z
\\
\omega_{xz}S_x-\omega_{yz}S_y-S_z \ea\,\, ,
\label{rotation}
\eea
we find (to leading order in $\omega_{\alpha\beta}$) that the easy axis term, $-AS_z^2$, is transformed into
\be
A(\omega_{xz}\{S_x,S_z\}+\omega_{yz}\{S_y,S_z\}).
\ee
Comparison with the last term in Eq.~(\ref{H_sp}) then  yields $g_4=2A$.

If the rotation matrices $R_\alpha$, $\alpha=x,y,z$, are expanded up to second order, one finds terms that include symmetric elements of the strain tensor $\epsilon$,
\be
R_x=\baaa 1 & 0 &
0 \\ 0 & 1-\frac{1}{2}\delta\phi_x^2 & -\delta\phi_x \\ 0 &
\delta\phi_x &
1-\frac{1}{2}\delta\phi_x^2 \ea,
\ee
\be
R_y=\baaa
1-\frac{1}{2}\delta\phi_y^2 & 0 & -\delta\phi_y \\ 0 & 1 & 0
\\
\delta\phi_y & 0 & 1-\frac{1}{2}\delta\phi_y^2 \ea,
\ee
\be
R_z=\baaa
1-\frac{1}{2}\delta\phi_z^2 & -\delta\phi_z & 0 \\ \delta\phi_z &
1-
\frac{1}{2}\delta\phi_z^2 & 0 \\ 0 & 0 & 1 \ea.
\ee
Now we obtain from $\bu=R_zR_yR_x\bx-\bx$
\be
\bu=\delta\bphi\times\bx-\frac{1}{2}\ba
\left(\delta\phi_y^2+
\delta\phi_z^2\right)x \\
\left(\delta\phi_x^2+\delta\phi_z^2\right)y
\\
\left(\delta\phi_x^2+\delta\phi_y^2\right)z \ea.
\ee
By keeping derivatives of $\delta\phi_\alpha$, up to second order we find
$
\delta\phi_x^2=\varepsilon_{xx}-\varepsilon_{yy}-\varepsilon_{zz},
$
and cyclic permutation of (x,y,z). \\
After inserting the rotated spin vector $R_xR_y\bS$ into
$
-AS_z^2=-A(\bS^2-S_x^2-S_y^2)
$
we get for the right-hand side
\be
A\left(\epsilon_{xx}-\epsilon_{yy}\right)\left(S_x^2-S_y^2\right)+O(\epsilon^2),
\ee
where we retain only terms that induce spin transitions.
Comparing with the spin-phonon Hamiltonian (\ref{H_sp}) one sees immediately that $g_1=A$, and thus 
\be
g_1=g_4/2=A.
\label{g} 
\ee
Thus the coupling constants $g_1$ and $g_4$ are explicitly expressed in terms of the anisotropy $A$.

Finally, we note that the terms in Eq.~(\ref{H_sp}) that are proportional to $g_{1,2}$ produce second-order transitions with $\Delta m=\pm 2$, while the ones proportional to $g_{3,4}$ produce first-order transitions with $\Delta m=\pm 1$. Thus, Eq.~(\ref{g}) implies that first-order and second-order transitions are equally important for the relaxation. In following Abragam and Bleaney,\cite{Abragam} it is now very plausible to adopt the approximations $|g_2|\approx g_1=A$ and $|g_3| \approx g_4=2A$ (the sign is irrelevant for the transition rates calculated below).


\section{Master equation including spin tunneling}
\label{master}

\subsection{Generalized master equation}
\label{generalized}

In this section we derive a master equation that describes the  relaxation of the spin due to phonon-assisted transitions including resonances due to tunneling.
For this we make use of a standard formalism\cite{Blum,Fick_Sauermann} suitable to describe a system (spin) coupled to a heat bath reservoir (phonons), the latter of which is in thermodynamic equilibrium described by the canonical density matrix $\rho_{\rm ph}$ for free phonons. That means we start from the full Hamiltonian $\cH=\cH_0+\cH_{\rm ph}+\cH_{\rm sp}$, where $\cH_0=\cH_{\rm a}+\cH_{\rm Z}+\cH_{\rm T}$ represents the system, $\cH_{\rm ph}$ the phonon heat bath, and $\cH_{\rm sp}$ given in Eq.~(\ref{H_sp}) is of the form $\cH_{\rm sp}=\sum_i Q_i\otimes F_i$, where $Q_i$ is a spin operator and $F_i$ is a phonon operator.
The generalized master equation in the interaction picture $(I)$ in Born and Markoff approximation reads [see Eq.~(8.1.22) in Ref.~\onlinecite{Blum}]
\bea
\dot{\rho}^I(t) & = &
-\left(\frac{1}{\hbar}\right)^2\sum\limits_{ij}\int_0^\infty dt''\nn\\
& &
\times\left\{\left[Q_i(t),Q_j(t-t'')\rho^I(t)\right]\left<F_i(t'')F_j\right>\right. \nn\\
& &
-\left.\left[Q_i(t),\rho^I(t)Q_j(t-t'')\right]\left<F_jF_i(t'')\right>
\right\},
\label{GME_interaction}
\eea
where $Q_i(t)=e^{ i\cH_0 t}Q_ie^{- i\cH_0 t}$.
Equation (\ref{GME_interaction}) is valid for the situations where the correlation time $\tau_c$ in the heat bath is much smaller than the relaxation time $\tau$ of the spin system. Indeed, the assumption is satisfied here since a rough estimate for thermal phonons yields $\tau_c\sim \hbar/k_B T\sim 10^{-11}$ s, at $T=1$ K, whereas it will turn out that $\tau\gtrsim 1$ s (see below). In this case, the integral kernel gives a vanishing contribution for times $t''$ larger than $\tau_c$, and thus one can extend the upper limit of the time integral to infinity and replace $\rho(t-t'')$ by $\rho(t)$ (see also Ref.~\onlinecite{Fick_Sauermann}, Chap.~13).

As our undamped Hamiltonian $\cH_0$ has also non-diagonal elements in the $\left|m\right>$-basis it proves convenient to formulate the generalized master equation in the Schr\"odinger picture, i.e., with
$\rho^I(t)=e^{(i/\hbar)\cH_0 t}\rho(t)e^{-(i/\hbar)\cH_0 t}$,
we get
\bea
\dot{\rho}(t) & = &
\frac{ i}{\hbar}\left[\rho(t),\cH_0\right]
-\left(\frac{1}{\hbar}\right)^2\sum\limits_{ij}\int_0^\infty dt''\nn\\
& &
\times\left\{\left[Q_i,Q_j(-t'')\rho(t)\right]\left<F_i(t'')F_j\right>\right. \nn\\
& &
-\left.\left[Q_i,\rho(t)Q_j(-t'')\right]\left<F_jF_i(t'')\right>
\right\}.
\label{Schroedinger}
\eea
As the tunnel splitting generated by $\cH_{\rm T}$ is smaller than the level spacing of $\cH_0$, i.e., $E_{mm'}<\left|\varepsilon_m-\varepsilon_{m'}\right|$ (see below), we can approximate the free propagator $e^{-(i/\hbar)\cH_0 t''}$ within the integral kernel by $e^{-(i/\hbar)(\cH_{\rm a}+\cH_{\rm Z}) t''}$ in the rest of our calculations.\cite{propagator}
Next, we take the matrix elements of Eq.~(\ref{Schroedinger}) using  $\rho_{mm'}=\left<m\left|\rho\right|m'\right>$, $\rho_m=\rho_{mm}$, $\cU_{mm'}=e^{-\frac{ i}{\hbar}(\varepsilon_m-\varepsilon_{m'})t''}$, and with the definitions\cite{Blum,Blum_analogy}
\bea
\Gamma_{mkln}^+ & = &
\frac{1}{\hbar^2}\sum\limits_{i,j}\left<m\left|Q_i\right|k\right>\left<l\left|
Q_j\right|n\right>\int_0^\infty dt''\cU_{ln}\left<F_i(t'')F_j\right>, \nn\\
\Gamma_{mkln}^- & = &
\frac{1}{\hbar^2}\sum\limits_{i,j}\left<m\left|Q_i\right|k\right>\left<l\left|
Q_j\right|n\right>\int_0^\infty dt''\cU_{mk}\left<F_jF_i(t'')\right>, \nn\\
\gamma_{m'm} & = &
\sum\limits_k
\left(\Gamma_{m'kkm'}^++\Gamma_{mkkm}^-\right)-\Gamma_{mmm'm'}^+-\Gamma_{mmm'm'}^-, \nn\\
W_{mn} & = & \Gamma_{nmmn}^++\Gamma_{nmmn}^-,
\label{Fermi}
\eea
it follows from Eq.~(\ref{Schroedinger}) that
\be
\dot{\rho}_{mm'}=\frac{ i}{\hbar}\left[\rho,\cH_0\right]_{mm'}
+\delta_{mm'}\sum\limits_{n\ne m}\rho_nW_{mn}-\gamma_{mm'}\rho_{mm'},
\label{GME}
\ee
where we have considered only the secular terms [i.e., the ``coarse-grained" derivative was taken with respect to $t$ in Eq.~(\ref{GME_interaction})]\cite{Blum} and set $\left[,\right]_{mm'}=\left<m\left|\left[,\right]\right|m'\right>$. The relaxation of the magnetization is entirely based on Eq.~(\ref{GME}).

The difference to the usual master equation is that Eq.~(\ref{GME}) takes also off-diagonal elements of the density matrix $\rho(t)$ into account. This is essential to describe tunneling of the magnetization, which is caused by the overlap of the $S_z$ states.

The diagonal elements ($m=m'$) of Eq.~(\ref{GME}) yield the master equation
\be
\dot{\rho}_m=\frac{ i}{\hbar}\left[\rho,\cH_0\right]_{mm}
+\sum\limits_{n\ne m}\rho_nW_{mn}-\rho_m\sum\limits_{n\ne m}W_{nm}.
\ee
The equation for the off-diagonal elements ($m\ne m'$),
\be
\dot{\rho}_{mm'}=\frac{ i}{\hbar}\left[\rho,\cH_0\right]_{mm'}-
\gamma_{mm'}\rho_{mm'},
\ee
can be simplified in the following way. According to Eq.~(\ref{Spin-phonon}), $Q_i$ is an element of the set $\left\{S_+^2,S_-^2,S_+S_z,S_zS_+,S_-S_z,S_zS_-\right\}$. Hence, we see that $\Gamma_{mmm'm'}^+=\Gamma_{mmm'm'}^-=0$, and we get
\bea
\gamma_{m'm}=\gamma_{mm'} & = & \frac{1}{2}\sum\limits_n\left(W_{nm'}+W_{nm}\right) \nn\\
& = & \frac{1}{2}\left(W_m+W_{m'}\right),
\eea 
where we use the abbreviation $W_m=\sum_nW_{nm}$.

Evaluation of Eq.~(\ref{Fermi}) leads immediately to Fermi's golden rule for transition rates in first quantization [see Eq.~(8.2.3) in Ref.~\onlinecite{Blum}]:
\bea
W_{mn} & = & \frac{2\pi}{\hbar}\sum\limits_{NN'}\left|\left<mN\left|H_{\rm sp}\right|nN'\right>\right|^2
\left<N'\left|\rho_{\rm ph}\right|N'\right> \nn\\
& & \times\left.\delta\left(E_{N'}-E_N-\varepsilon_m+\varepsilon_n\right)\right. .
\label{FGR}
\eea
Explicit evaluation yields (see Appendix \ref{sp}),
\bea
W_{m\pm 1,m} & = & \frac{A^2s_{\pm 1}}{12\pi\rho c^5\hbar^4}
\frac{(\varepsilon_{m\pm 1}-\varepsilon_m)^3}
{e^{\beta(\varepsilon_{m\pm 1}-\varepsilon_m)}-1}\, ,
\nn\\
W_{m\pm 2,m} & = & \frac{17A^2s_{\pm 2}}{192\pi\rho c^5\hbar^4}
\frac{(\varepsilon_{m\pm 2}-\varepsilon_m)^3}
{e^{\beta(\varepsilon_{m\pm 2}-\varepsilon_m)}-1}\,
,
\label{sp_rates}
\eea
where $s_{\pm 1}=(s\mp m)(s\pm m+1)(2m\pm 1)^2$, and $s_{\pm 2}=(s\mp m)(s\pm m+1)(s\mp m-1)(s\pm m+2)$. The mass density $\rho$ for Mn$_{12}$ is given by $1.83\times 10^3$ kg/m$^3$.\cite{Lis} Here, $c$ is the sound velocity of the Mn crystal, which is the only free parameter in our theory. As already mentioned, we are not aware of direct measurements of $c$ (but see below). Note that the transition rates $W_{m\pm 1,m}, W_{m\pm 2,m}$ are very sensitive to variations of the sound velocity $c$, as the latter enters with the fifth power.


\subsection{Spin tunneling}
\label{spin}

We include now the spin tunneling in the generalized master equation (\ref{GME}). Let $\left|m\right>$ and $\left|m'\right>$ be two eigenstates of $\cH_{\rm a}+\cH_{\rm Z}$ on the left and right sides of the barrier, respectively. $\left|m\right>$ and $\left|m'\right>$ are degenerate when $\delta H_z=H_z^{mm'}-H_z$ vanishes. In the presence of tunneling, induced by ${\cal H}_{\rm T}$, the two states form (anti)symmetric levels split by $E_{mm'}$ (for  $\delta H_z=0$). By using time-independent perturbation theory in higher order the tunnel splitting energy $E_{mm'}$ can be evaluated,\cite{Garanin}
\be
E_{mm'}=2\frac{V_{m,m-1}}{\varepsilon_{m-1}-\varepsilon_m}\frac{V_{m-1,m-2}}
{\varepsilon_{m-2}-\varepsilon_m}\cdots V_{-m'+1,-m'}.
\label{splitting}
\ee
Note that in this expression only steps with $\Delta m=\pm1$ are allowed. However, for the present purpose we need to generalize Eq.~(\ref{splitting}) to situations where potentials $V_{m_i,m_{i+1}}\in \Bbb{R}$ with arbitrary steps $\Delta m=m_i-m_{i+1}$ ($m>m_i>m_{i+1}>m'$, $i=1,\ldots,N-1$) can occur. As we will show in Appendix \ref{resolvent} this is indeed possible by using resolvent techniques, and we find
\be
E_{mm'}=2\left|\sum\limits_{m_1,\ldots,m_N \atop m_i\ne m,m'}
\frac{V_{m,m_1}}{\varepsilon_m-\varepsilon_{m_1}}\prod\limits_{i=1}^{N-1}
\frac{V_{m_i,m_{i+1}}}{\varepsilon_m-\varepsilon_{m_{i+1}}}V_{m_N,m'}\right|,
\label{gsplitting}
\ee
where $N$ is the lowest order of the degenerate perturbation theory, by means of which Eq.~(\ref{gsplitting}) has been derived ($N$th-order secular equation\cite{Landau}), giving a nonvanishing contribution to $E_{mm'}$. For the potentials $V_{m_i,m_{i+1}}$ we insert combinations of terms occurring in ${\cal H}_{\rm T}$. For example, the anisotropy $B_4$ leads to transitions $\Delta m=\pm 4$, while the misalignment $H_x$ leads to transitions $\Delta m=\pm 1$. The summations in Eq.~(\ref{gsplitting}) can be thought of as summation over different paths in the Hilbert space connecting $\left|m\right>$ with $\left|m'\right>$.  

Continuing the evaluation of the first part of Eq.~(\ref{GME}) we project the undamped Hamiltonian $\cH_0$ by $P=\sum_{n=m,m'}\left|n\right>\left<n\right|$ on the two-state system $\left\{\left|m\right>,\left|m'\right>\right\}$, which yields the two-state Hamiltonian in the presence of a bias field (see Fig.~\ref{tunnel-fig})
\bea
{\overline{{\cal H}}}_{\rm T} & = & \xi_m\left|m\right>\left<m\right|
+\frac{E_{mm'}}{2}\left|m\right>\left<m'\right|\,\,+ (m \leftrightarrow m') \\
& \hat{=} & \left[\begin{array}{cc} \xi_m
& \frac{E_{mm'}}{2} \\ \frac{E_{mm'}}{2} & \xi_{m'}
\end{array}\right]=P\cH_0P, \nn
\eea
with $\xi_m=\varepsilon_m+g\mu_B \delta H_z m$ and the energy eigenvalues $E_{\rm T}=\frac{1}{2}\left[\xi_m+\xi_{m'}\pm\sqrt{\left(\xi_m-\xi_{m'}\right)^2+E_{mm '}^2}\right]$. ${\overline{{\cal H}}}_{\rm T}$ provides a valid description as long as the level splitting remains smaller than the level spacing, i.e.,
\be
\Delta=\sqrt{(\xi_m-\xi_{m'})^2+E_{mm'}^2}\ll |\varepsilon_{m^({'}^)}-\varepsilon_{m^({'}^)\pm 1}|.
\label{Delta}
\ee 
We have checked that between two main resonances this condition is satisfied for the states $\left|m_{\rm T}\right>$ and $\left|m_{\rm T}'\right>$ of the dominant paths and for each degenerate pair of states $\left|m\right>$, $\left|m'\right>$ with $\varepsilon_m,\varepsilon_{m'}\le\varepsilon_{m_{\rm T}},\varepsilon_{m_{\rm T}'}$ (see Sec.~\ref{paths}).

Next we insert the two-state Hamiltonian ${\overline{{\cal H}}}_{\rm T}$ into the generalized master equation (\ref{GME}), which yields
\be
\dot{\rho}_m=\frac{ i E_{mm'}}{2{\hbar}}\left(\rho_{mm'}-\rho_{m'm}\right)-
W_m\rho_m+\!\!\!\sum_{n\ne m,m'}\!\!\!W_{mn}\rho_n
\label{rho}
\ee
and
\bea
\dot{\rho}_{mm'}\!
=\!-\!\left(\!\frac{ i}{\hbar}\xi_{mm'}+\gamma_{mm'}\!\right)\!\rho_{mm'}\!
+\frac{ i E_{mm'}}{2\hbar}\!\left(\rho_m\!-\!\rho_{m'}\right),
\label{rho1}
\eea
where $\xi_{mm'}=\xi_{m}-\xi_{m'}$, and likewise for $m\leftrightarrow m'$. 
Ultimately, we are interested in the overall relaxation time $\tau$ of the quantity $\rho_s-\rho_{-s}$ (see Sec.~\ref{paths}) due to phonon-induced transitions. This $\tau$ turns out to be much longer than $\tau_{\rm d}=1/\gamma_{mm'}$, which is the decoherence time of the decay of the off-diagonal elements $\rho_{mm'}\propto e^{-t/\tau_{\rm d}}$ of the density matrix $\rho$. Thus, we can neglect the time-dependence of the off-diagonal elements, i.e., $\dot{\rho}_{mm'}\approx 0$. Physically this means that we deal with incoherent tunneling for times $t>\tau_{\rm d}$.\cite{coherence} Inserting then the stationary solution of Eq.~(\ref{rho1}) into Eq.~(\ref{rho}), which leads to the complete master equation including resonant as well as nonresonant levels,
\be
\dot{\rho}_m =-W_m\rho_m+\!\!\!\sum_{n\ne m,m'}\!\!\!
W_{mn}\rho_n\,+\Gamma_m^{m'}\left(\rho_{m'}-\rho_m\right),
\label{finalmeq}
\ee
where
\be
\Gamma_m^{m'}=E_{mm'}^2\frac{W_m+W_{m'}}{4\xi_{mm'}^2
+\hbar^2\left(W_m+W_{m'}\right)^2}
\label{Lorentzian}
\ee
is the transition rate from $m$ to $m'$ (induced by tunneling) in the presence of phonon damping.\cite{Fort_formula} The relaxation dynamics of the resonances described by $1/\Gamma_m^{m'}\sim 10^{-7}$ s (see Fig.~\ref{truncation}) turns out to be much faster than the phonon-induced overall relaxation, i.e., $1/\Gamma_m^{m'}\ll\tau\gtrsim 1$s (see Fig.~\ref{overall}). 
Thus, our derivation based on the assumption $\tau\gg\tau_{\rm d}$ is self-consistent since $1/\Gamma_m^{m'}\sim\tau_{\rm d}$. 
Note that Eq.~(\ref{finalmeq}) is now of the usual form of a master equation, i.e., only  diagonal elements of the density matrix $\rho(t)$ occur. For levels $k\neq m,m'$, Eq.~(\ref{finalmeq}) reduces to
\be
\dot{\rho}_k=-W_k\rho_k+\sum_nW_{kn}\rho_n.
\ee

We note that  $\Gamma_m^{m'}$ has a Lorentzian shape with respect to the external magnetic field $\delta H_z$ occurring in $\xi_{mm'}$. The  $H_z$ dependence of $W_m$ around the resonances turns out to be much weaker (see below) and can be safely ignored. It is thus this $\Gamma_m^{m'}$ that will determine the peak shape of the magnetization resonances (see below and Figs.~\ref{overall}--\ref{fourpeaks}). Note that in
Figs.~\ref{overall}--\ref{fourpeaks}
these Lorentzians are {\it truncated} at the center of the peak by the spin-phonon transition rates $W_m$ and $W_{m'}$ in such a way that the effective linewidth (defined as the width at half of the height of the truncated peak) is much larger than $(W_m+W_{m'})/2$. This needs some further explanations, which are given in Sec.~\ref{lorentzian}, after we have discussed the relaxation times. 

\begin{figure}[htb]
  \begin{center}
\leavevmode
\epsfxsize=8.5cm
\epsffile{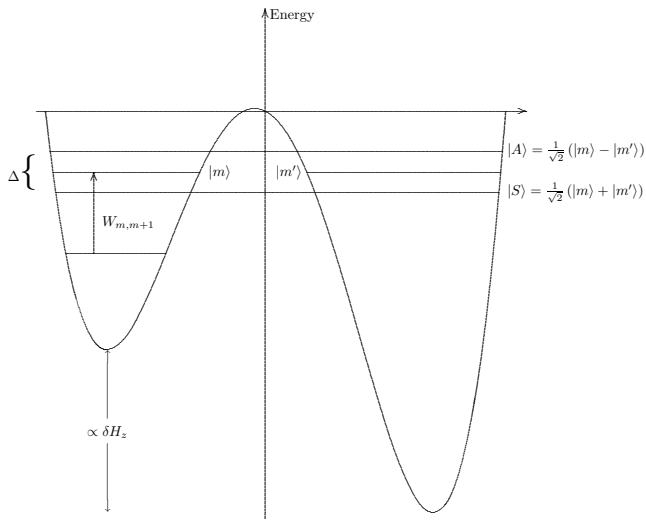}
\end{center}
\caption{Tunneling configuration.}
\label{tunnel-fig}
\end{figure}


\section{Relaxation time}
\label{relax}

\subsection{Numerical diagonalization of the master equation}

In this section we give the results of our exact evaluation obtained by a numerical diagonalization of the master equation. For convenience we now write down the master equation (\ref{finalmeq}) as a vector equation,
\be
\dot{\vec {\rho}}(t)={\tilde W} {\vec \rho}(t),
\label{veceq}
\ee
where the elements of the vector ${\vec \rho}(t)$ are the diagonal elements of the density matrix $\rho$. Within the interval $I_{n_1,n_2}$ [see Eq.~(\ref{interval})] delimited by the two main resonances $n_1$ and $n_2$ only the tunneling rates $\Gamma_{m_1}^{m'_1}$ and  $\Gamma_{m_2}^{m'_2}$ [Eq.~(\ref{Lorentzian})] for which $m_1+m'_1=n_1$ and $m_2+m'_2=n_2$ (see Sec.~\ref{analytical}) are allowed to be included for self-consistency reasons: the tunnel splitting (\ref{Delta}) entering $\Gamma$ is only valid within this interval $I_{n_1,n_2}$. If $w_i$, $i=1,2,\ldots,21$, are the eigenvalues of the rate matrix ${\tilde W}$, the dominant relaxation time of the spin system is given by
\be
\tau=\max_i\left\{-\frac{1}{\text{Re }w_i}\right\}.
\ee
The eigenvalues $w_i$ turn out to be nondegenerate with the smallest one being far separated by a factor of at least $10^4$ from the remaining ones. The result is plotted in Fig.~\ref{overall}, where the overall relaxation rate $\tau$ is shown as a function of $H_z$ at $T=1.9$ K. It is important to note that all the resonance peaks are of Lorentzian shape. We note that in our model the even resonances are induced by the quartic $B_4$ anisotropy, whereas the odd resonances are induced by product combinations of $B_4S_\pm^4$ and $H_x S_x$ terms [see Eq.~(\ref{gsplitting})]. For the plot in Fig.~\ref{overall} we set $\theta=1^\circ$ in accordance with the experimental uncertainty,\cite{FRIEDMAN} leading to a maximal transversal field $H_x$ of about 350 G.

\begin{figure}
  \begin{center}
\leavevmode
\epsfxsize=8.5cm
\epsffile{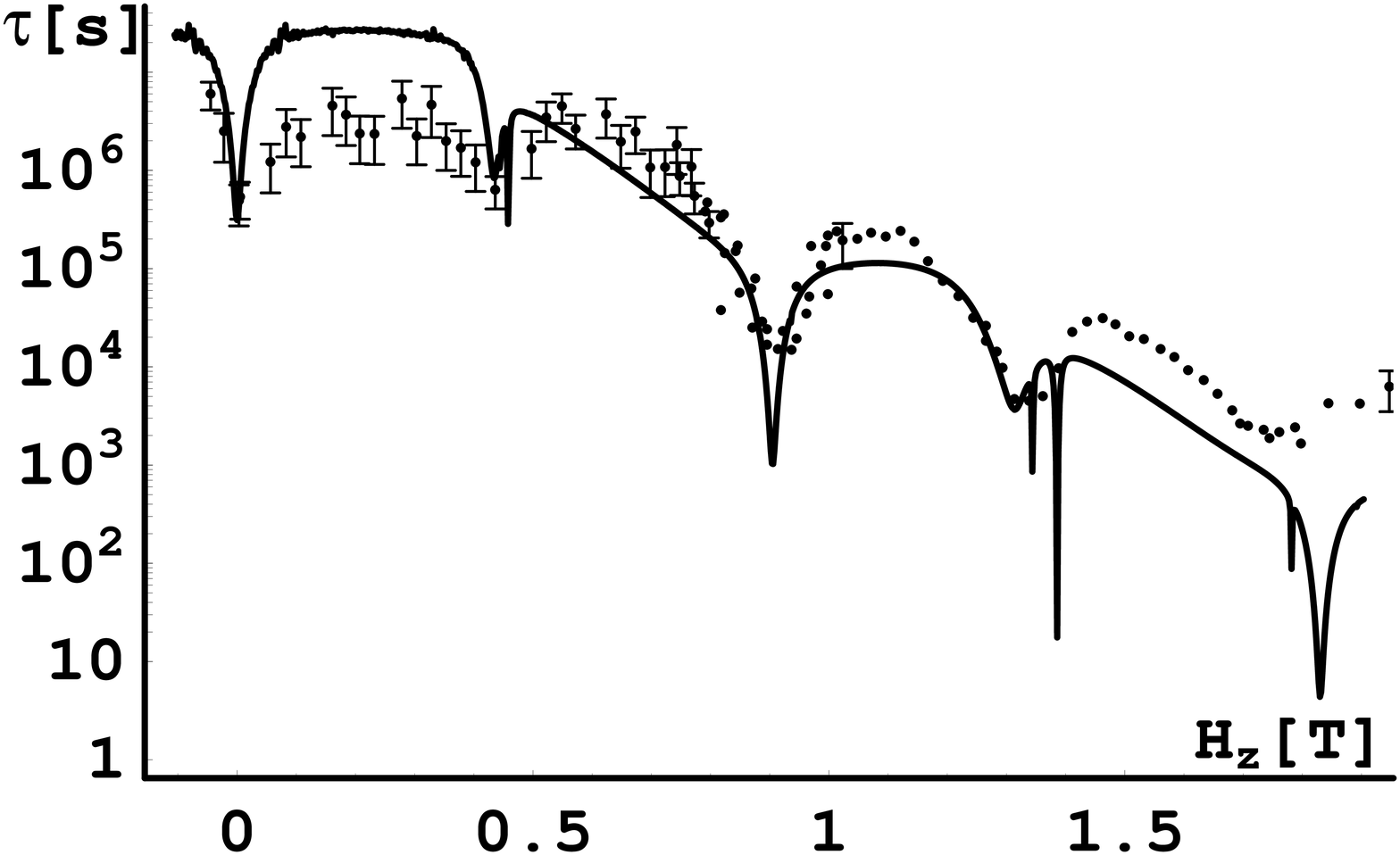}
\end{center}
\caption{Full line: semilogarithmic plot of calculated relaxation time $\tau$ as function of  magnetic field $H_z$ at $T=1.9$ K. The optimal fit values (see text) are $A/k_B=0.54$ K, $B/k_B=1.1\times 10^{-3}$ K, and $B_4/k_B=8.5\times 10^{-5}$ K, $\theta=1^\circ$, and $c=1.45\times 10^3$ m/s. Dots and error bars: data taken from Ref.~\protect\onlinecite{Thomas}.}
\label{overall}
\end{figure}

\subsection{Comparison with experimental data}

For comparison we also include in Fig.~\ref{overall} the data reported by Thomas {\it et al.}\cite{Thomas,angle} We have optimized the fit  (as explained in the Introduction) in such a way that the fits of the model parameters, given by
\bea
A/k_B & = & 0.54\text{ K}, \label{A}\\
B/k_B & = & 1.1\times 10^{-3}\text{ K}, \label{B}\\
B_4/k_B & = & 8.5\times 10^{-5}\text{ K}, \label{B4}
\eea
are roughly within the reported experimental uncertainties of Refs.~\onlinecite{Barra} and \onlinecite{Zhong} (see above). 
The value of $B_4$ is in excellent agreement with recent measurements performed in Ref.~\onlinecite{Gudel}.
Our fit of the sound velocity yields
\be
c=1.45\times 10^3\text{ m/s}.
\label{c_fit1}
\ee
There is a difference between odd and even resonances, i.e., the relaxation time $\tau$ at an even resonance peak is about three times smaller than the one at a subsequent odd resonance peak.  It should be mentioned that almost identical plots are obtained for $0.5^\circ\lesssim\theta\lesssim 3^\circ$, as can be seen in Figs.~\ref{overall}--\ref{overall3}.  
The present theory holds for  $\left | H_x\right |\lesssim 1000$ G (which is well satisfied here), otherwise the shift of the  levels $\left |m\right>$ due to the perturbation $H_x S_x$ must be taken into account. For example, for the resonance $n$=3 the relevant tunneling takes place between $\left|4\right>$ and $\left|-1\right>$. The dominant second-order shift
\bea
\frac{\left<1\left|g\mu_BH_xS_x\right|0\right>\left<0\left|g\mu_BH_xS_x\right|1 \right>}{\varepsilon_1-\varepsilon_0} & &  \nn\\
+\frac{\left<1\left|g\mu_BH_xS_x\right|2\right> \left<2\left|g\mu_BH_xS_x\right|1\right>}{\varepsilon_1-\varepsilon_2} & = & k_B\times 40\text{ mK} \nn\\
& \ll & \left|\varepsilon_1\right|=k_B\times 2.3\text{ K} 
\eea
clearly shows that the unperturbed states $\left\{\left|m\right>\right\}$ are a good zeroth-order approximation. It is also important to know whether the second-order shifts caused by the perturbation $\cH_{\rm T}$ are negligible compared to the tunnel splitting $E_{m_{\rm T}m_{\rm T}'}$. Explicitly, we find
\bea
\left|\Delta_2^{(2)}-\Delta_{-6}^{(2)}\right|/k_B & = &
8.5\text{ mK} \quad (n=4),\nn\\
\left|\Delta_1^{(2)}-\Delta_{-4}^{(2)}\right|/k_B & = &
13.2\text{ mK} \quad (n=3),\nn\\
\left|\Delta_3^{(2)}-\Delta_{-5}^{(2)}\right|/k_B & = &
5.0\text{ mK} \quad (n=2),\nn\\
\left|\Delta_2^{(2)}-\Delta_{-3}^{(2)}\right|/k_B & = &
0.5\text{ mK} \quad (n=1),\nn\\
\left|\Delta_4^{(2)}-\Delta_{-4}^{(2)}\right|/k_B & = &
0 \quad (n=0),
\eea
where $\Delta_{m_{\rm T}}^{(2)}$ is the second-order shift of the unperturbed states $\left|m_{\rm T}\right>$ of the dominant paths. These renormalizations cause a very small shift of the resonance peaks, e.g., the shift for $n=3$ is 0.6 mT. The relevant tunnel splitting energies of the odd and even resonances are about the same (except $E_{4,-1}$): 
\bea
E_{4,-4} & \approx & E_{3,-2} \approx E_{5,-3} \approx k_B\times 45\text{ mK, } \nn\\
E_{4,-1}/k_B & \approx & 130\text{ mK,  }
E_{6,-2}/k_B \approx 40\text{ mK}.
\label{E}
\eea
For comparison, $E_{2,-2}/k_B\approx 1$ K [see also Eq.~(\ref{dominant_channel})]. In conclusion, the diagonal elements of the shifts of the nondegenerate perturbation theory are much smaller than the off-diagonal elements of the shifts of the degenerate perturbation theory (see Appendix \ref{resolvent}). Thus our assumption of quasidegeneracy is very well satisfied.

\begin{figure}
  \begin{center}
\leavevmode
\epsfxsize=8.5cm
\epsffile{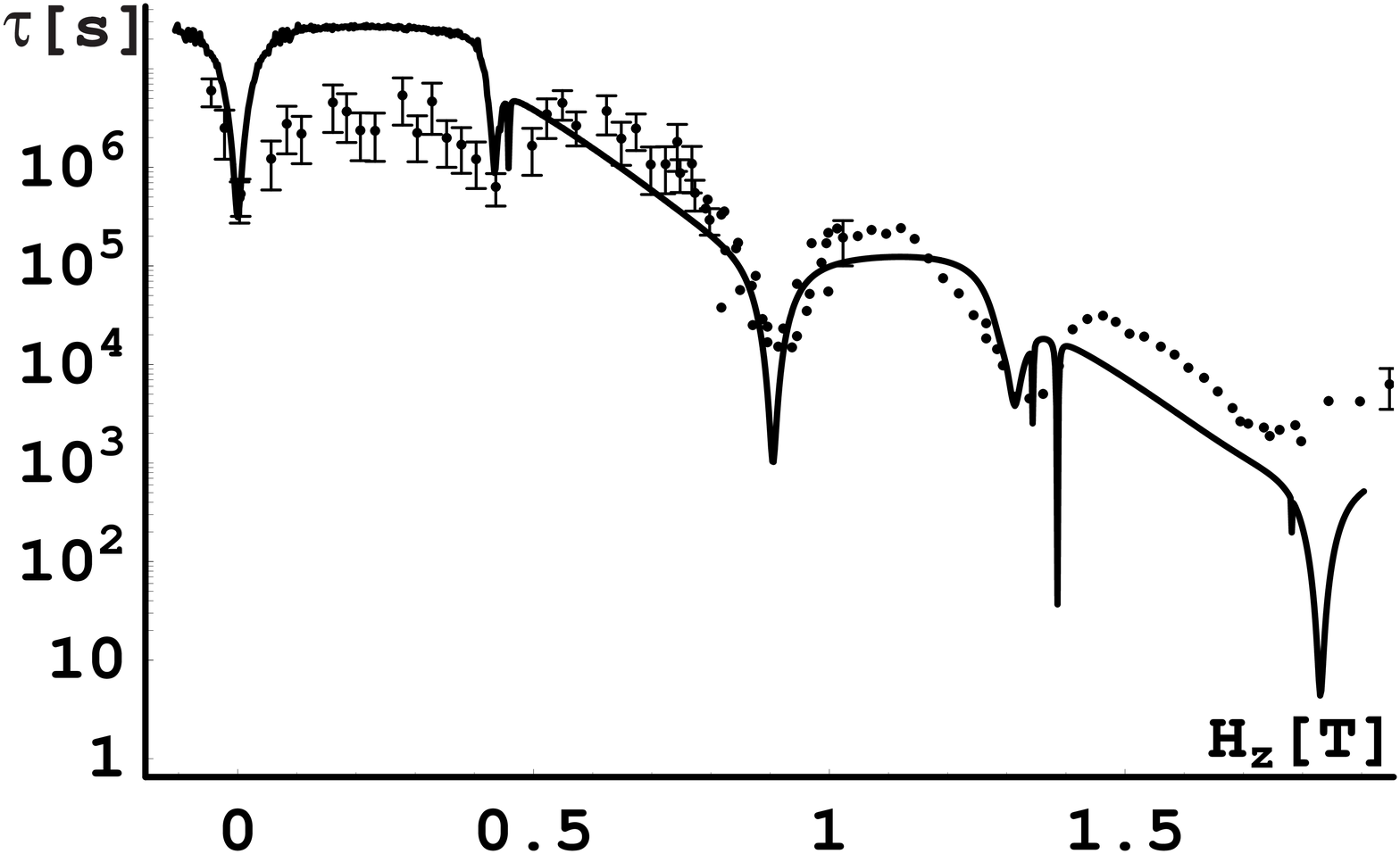}
\end{center}
\caption{Full line: semilogarithmic plot of calculated relaxation time $\tau$ as function of  magnetic field $H_z$ at $T=1.9$ K. Here $\theta=0.5^\circ$ has been chosen. Dots and error bars: data taken from Ref.~\protect\onlinecite{Thomas}.}
\label{overall05}
\end{figure}

\begin{figure}
  \begin{center}
    \leavevmode
\epsfxsize=8.5cm
\epsffile{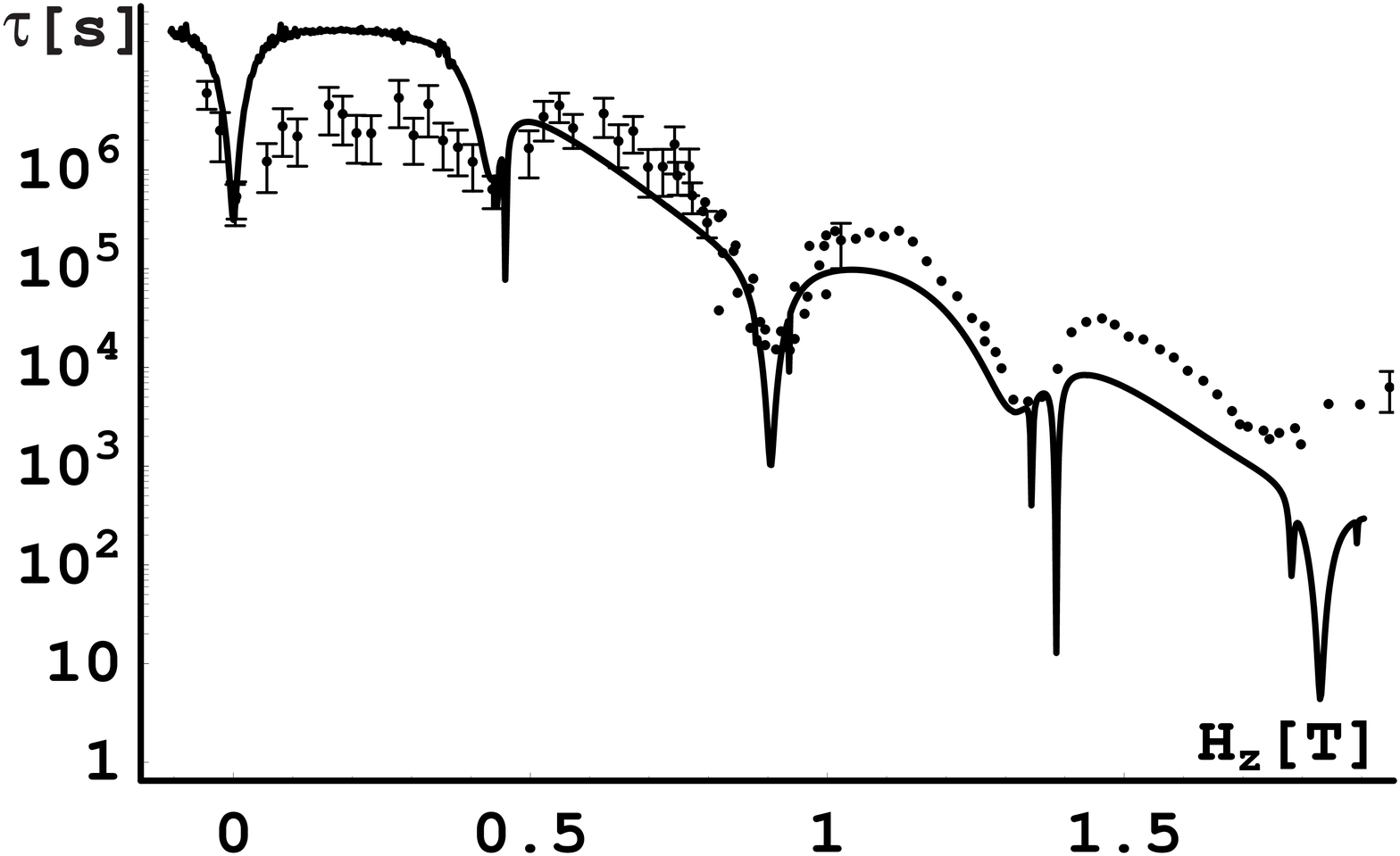}
\end{center}
\caption{Full line: semilogarithmic plot of calculated relaxation time $\tau$ as function of  magnetic field $H_z$ at $T=1.9$ K. Here $\theta=2^\circ$ has been chosen. Dots and error bars: data taken from Ref.~\protect\onlinecite{Thomas}.}
\label{overall2}
\end{figure}

\begin{figure}
  \begin{center}
    \leavevmode
\epsfxsize=8.5cm
\epsffile{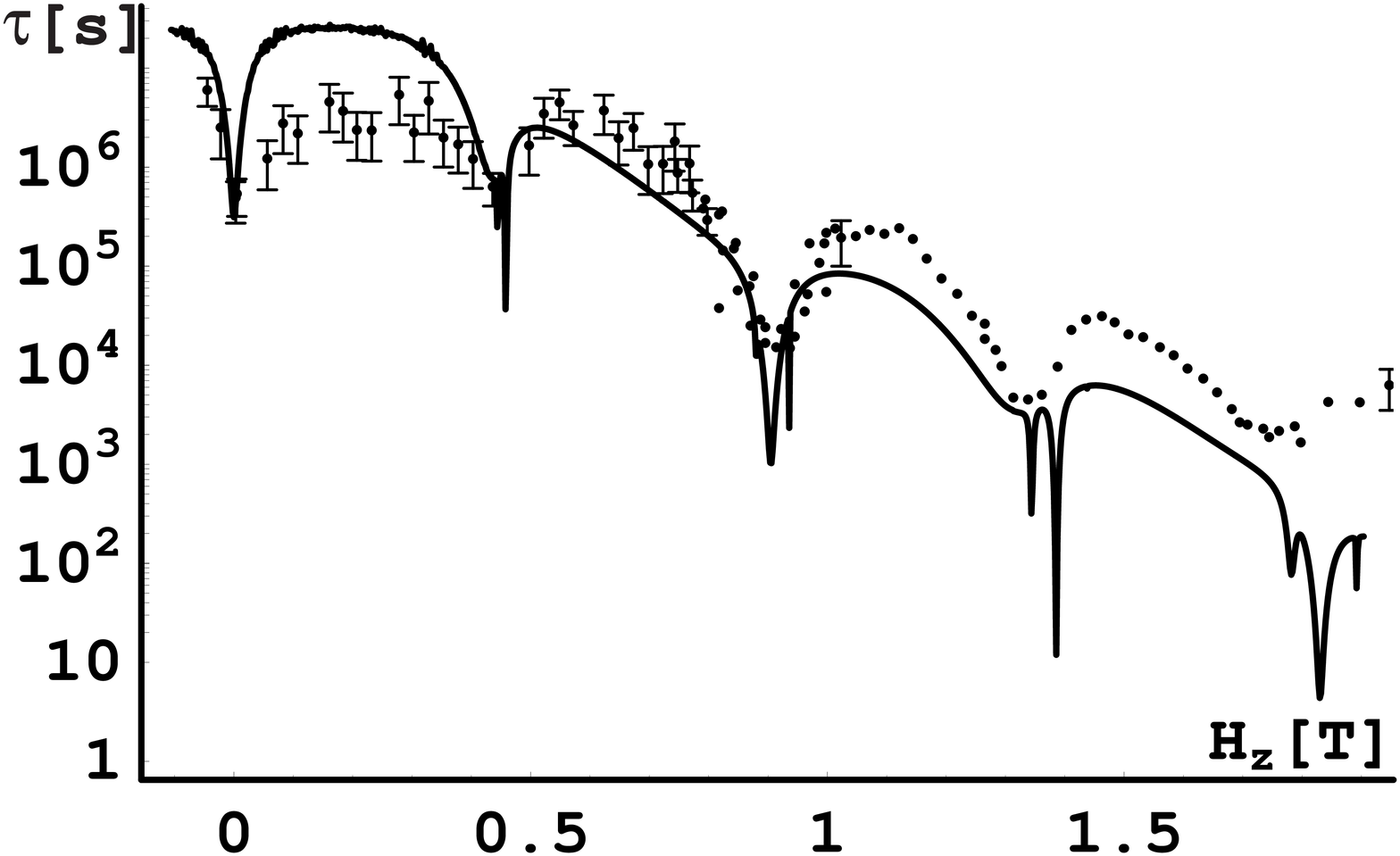}
\end{center}
\caption{Full line: semilogarithmic plot of calculated relaxation time $\tau$ as function of  magnetic field $H_z$ at $T=1.9$ K. Here $\theta=3^\circ$ has been chosen. Dots and error bars: data taken from Ref.~\protect\onlinecite{Thomas}.}
\label{overall3}
\end{figure}

We note that there are satellite peaks in Figs.~\ref{overall}--\ref{overall3}, the origin of which will be explained below in Sec.~\ref{paths}.

In Figs.~\ref{singlepeak} and \ref{fourpeaks} we plot the peaks of the first resonance at $H_z=0$, which is induced only by the $B_4$ anisotropy, for four different temperatures, namely $T=2.5,2.6,2.7$, and $2.8$ K. The four peaks (like all others) are of single Lorentzian shape as a result of the two-state transition rate $\Gamma_m^{m'}$ given in Eq.~(\ref{Lorentzian}). For comparison we plot in Figs.~\ref{singlepeak} and \ref{fourpeaks} the data reported by Friedman {\it et al.}\cite{FRIEDMAN} for the same temperatures (no error bars, however, are given in Ref.~\onlinecite{FRIEDMAN}). The optimal fit values are
\bea
A/k_B & = & 0.56\text{ K}, \\
B/k_B & = & 1.3\times 10^{-3}\text{ K}, \\
B_4/k_B & = & 14.4\times 10^{-5}\text{ K}, \\
c & = & 2.0\times 10^3\text{ m/s}.
\label{c_fit2}
\eea

\begin{figure}
  \begin{center}
    \leavevmode
\epsfxsize=8.5cm
\epsffile{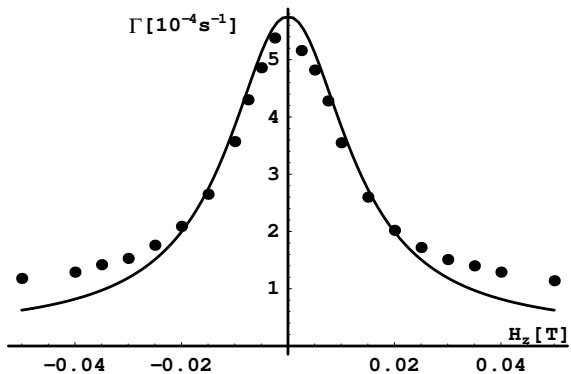}
\end{center}
\caption{ Full line: plot of calculated relaxation rate $\Gamma=1/\tau$ as function of $H_z$ for the first resonance peak at $T=2.6$ K. The Lorentzian shape is due to $\Gamma_m^{m'}$ in Eq.~(\protect\ref{Lorentzian}). The optimal fit values (see text) are $A/k_B=0.56$ K, $B/k_B=1.3\times 10^{-3}$ K, and $B_4/k_B=14.4\times 10^{-5}$ K, $\theta=1^\circ$, and  $c=2.0\times 10^3$ m/s. Dots: data taken from Ref.~\protect\onlinecite{FRIEDMAN}.}
\label{singlepeak}
\end{figure}

\begin{figure}
  \begin{center}
    \leavevmode
\epsfxsize=8.5cm
\epsffile{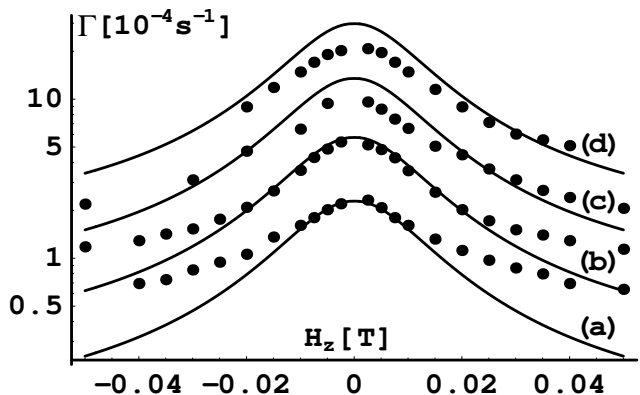}
\end{center}
\caption{ Full lines: semilogarithmic plots of calculated relaxation rate $\Gamma=1/\tau$ as function of $H_z$ for the first resonance peak at (a) $T=2.5$ K, (b) $T=2.6$ K, (c) $T=2.7$ K, and (d) $T=2.8$ K. All peaks are of single Lorentzian shape. The optimal fit values are the same as in Fig.~\protect\ref{singlepeak}. Dots: data taken from Ref.~\protect\onlinecite{FRIEDMAN}.}
\label{fourpeaks}
\end{figure}

Note that these values are the same for all four temperatures, which means that our peaks fit also the temperature dependence of the relaxation time. The fitting parameters turn out to be somewhat larger than the ones used in Fig.~\ref{overall} [see Eqs.~(\ref{A})--(\ref{B4})], which could be caused by sample differences, e.g., in volume-to-surface ratio and/or in shape anisotropy of the samples, etc. Indeed, the sample of Ref.~\onlinecite{FRIEDMAN} consists of many small crystallites in contrast to the single crystal used by Thomas {\it et al.}\cite{Thomas} In any case, the differences are small, and the sound velocity $c$ seems to be within the expected order of magnitude. Clearly, it would be highly desirable to check this prediction by an independent and direct measurement of $c$. On the other hand,\cite{Friedman_com} we can get an independent estimate for c from the specific heat and the Debye temperature $\Theta_D$ which was recently measured in Mn$_{12}$.\cite{Gomes} The reported value is $\Theta_D=(38\pm 4)$ K, and making use of the Debye relation\cite{Ashcroft}
\be
k_B\Theta_D=\hbar \omega_D=\hbar ck_D\text{, with } n=\frac{k_D^3}{6\pi^2}=\frac{1}{V_0},
\ee
we find
\be
c=(1.77-2.18)\times 10^3\text{ m/s},
\ee
where $\omega_D$ is the Debye frequency, $k_D$ the Debye wave vector, and $V_0=3716$ \AA$^3$ the unit-cell volume. Comparing this value for $c$ with the one obtained before, see Eqs.~(\ref{c_fit1}) and (\ref{c_fit2}), we see that the agreement is very good. This result corroborates not only our prediction of $c$ but also our values obtained for the spin-phonon coupling constants $g_i$.

Finally we also mention that the prefactor $A^2s_{\pm 1}/12\pi\rho c^5\hbar^4$
of our spin-phonon rates [Eq.~(\ref{sp_rates})] is in excellent agreement with
the value of the parameter denoted by
$C$ in a recent paper.\cite{Lascialfari} Note that their fit of the parameter\cite{Lascialfari} $C$ is not as precise as ours, because $C$ is assumed to be independent of the spin states $\left\{\left|m\right>\right\}$.

To summarize our results obtained so far, we see that the agreement between theory and experiment is satisfactory; in particular we emphasize that there is no free fit parameter. Thus, our model and its evaluation seems to contain the essential physics responsible for the magnetization relaxation in Mn$_{12}$.

\subsection{Comparison with previous results}

In comparison to previous results we obtain much better agreement between theory and experiment for the following reasons. For this comparison we can restrict ourselves to the work of Fort {\it et al.},\cite{Fort} since --- as far as we are aware of --- it has produced the best agreement with the relaxation data\cite{Thomas} thus far. First, the spin-phonon coupling constants $g_1$, $g_2$, and $g_3$ are explicitly given in our work for the first time ($g_4$ has been found before\cite{VILLAIN}). As shown in Sec.~\ref{model} we find them to be  of order $A=0.56$ K, and it is this value which leads to good agreement with all known experimental data which involve these coupling constants.\cite{Friedman,Thomas,Lascialfari} In contrast, Fort {\it et al.}\cite{Fort} set arbitrarily the $g_i$'s to values of 15 K and 30 K, which is clearly in contradiction to our microscopic values. Moreover, our value for the fit parameter $B_4$ fulfills the constraints of independent measurements,\cite{Fernandez,SESSOLI,Barra,Zhong} while Fort {\it et al.}\cite{Fort} obtain a $B_4$ value which is about 30 times smaller than the measured value in Ref.~\onlinecite{Barra}. 
From Figs.~\ref{singlepeak} and \ref{fourpeaks} we see that the temperature dependence of the relaxation time agrees quite well with the measurements of Friedman {\it et al.}\cite{FRIEDMAN}. Such a fit has not been attempted before.


\section{Relaxation paths}
\label{paths}

\subsection{Analytical result}
\label{analytical}

In order to get a better physical understanding for the relaxation process of the spin system it is instructive to determine the dominant transition paths via which the spin can relax into its ground state. For this we derive an approximate analytic expression for the relaxation time denoted by $\tau^*$ (to distinguish it from the exact $\tau$ obtained in the previous section). First, we solve the master equation for one {\it particular} transition path $n$ which does not intersect with other paths. For $H_z\ge 0$ we find (derivation is given below)
\be
\tau_n=\frac{1}{1+e^{\beta(\varepsilon_{-s}-\varepsilon_s)} }
\sum_{\{m_i\}_n}\frac{e^{\beta(\varepsilon_{m_i}-\varepsilon_s)}}{\Omega_{m_i}^{m_{i+1}}},
\label{analytic}
\ee
where $\Omega_{m_i}^{m_{i+1}}=\Omega_{m_i\to m_{i+1}}=W_{m_{i+1}m_i}$ or $\Gamma_{m_{i+1}}^{m_i}$, depending on the particular path $n$ characterized by the sum over the levels $m$ (see Figs.~\ref{diagram1} and \ref{diag1ser}). Equation (\ref{analytic}) holds for arbitrary initial ($\varepsilon_i$) and final ($\varepsilon_f$) energies, and for arbitrary steps $\Delta m=m_{i+1}-m_i$ (see below).

We now turn to the derivation of the relaxation time of a cascade including the external field $H_z$. For this we need to go beyond the results obtained previously\cite{Villain} for $H_z=0$, which requires a non-trivial extension. We start with rewriting the master equation (\ref{finalmeq}) as
\bea
\dot{\rho}_s & = & \Omega_{m_1}^s\rho_{m_1} -\Omega_s^{m_1}\rho_s, \nn\\
\dot{\rho}_{m_j} & = & \Omega_{m_{j+1}}^{m_j}\rho_{m_{j+1}}+\Omega_{m_{j-1}}^{m_j}\rho_{m_{j-1}} \nn\\
& & -\left.\Omega_{m_j}^{m_{j+1}}\rho_{m_j}-\Omega_{m_j}^{m_{j-1}}\rho_{m_j},\right.\nn\\
\dot{\rho}_{-s} & = & \Omega_{m_p}^{-s}\rho_{m_p}-\Omega_{-s}^{m_p}\rho_{-s},
\label{me}
\eea
with $m_j\in \left]-s,s\right[,\; m_j>m_{j+1},\; j=1,\ldots,p\le 2s-1$, and $\dot{\rho}_{m_j}=d\rho_{m_j}/dt$.
We consider now the stationary limit of Eq.~(\ref{me}) which we define by
\be
\dot{\rho}_s=-J,\;\dot{\rho}_{m_j}=0,\;\dot{\rho}_{-s}=J,
\label{stationary_limit}
\ee
where the first and last equation express conservation of the probability current $J$, which we assume positive for $H_z\ge 0$ and independent of $m$.
Equation (\ref{stationary_limit}) leads to $p+1$ equations,
\be
J(t)=\Omega_{m_i}^{m_{i+1}}\rho_{m_i}-\Omega_{m_{i+1}}^{m_i}\rho_{m_{i+1}},
\ee
and by solving for $\rho_{m_{i+1}}$ we get
\be
\rho_{m_i}=\frac{\Omega_{m_{i+1}}^{m_i}}{\Omega_{m_i}^{m_{i+1}}}\rho_{m_{i+1}}
+\frac{J}{\Omega_{m_i}^{m_{i+1}}},
\ee
where we have introduced $i=0,\ldots,p\le 2s-1$, and $m_0=s$.
To simplify the following treatment we assume detailed balance also for the tunneling processes. This approximation has little effect on the final result which turns out to agree very well with the exact relaxation time $\tau$ where no such approximations are invoked. Inserting then the detailed balance relation $\Omega_{m_{i+1}}^{m_i}/\Omega_{m_i}^{m_{i+1}}=e^{\beta(\varepsilon_{m_{i+1}}-\varepsilon_{m_i})}$ one obtains
\be
\rho_{m_i}=e^{\beta(\varepsilon_{m_{i+1}}-\varepsilon_{m_i})}\rho_{m_{i+1}} +\frac{J }{\Omega_{m_i}^{m_{i+1}}}.
\ee 
In order to get an equation that depends on $\rho_{-s}$ and $\rho_s$ only one has to sum over the following $p+1$ equations:
\bea
\rho_s & = &
e^{\beta(\varepsilon_{m_1}-\varepsilon_s)}\rho_{m_1}+\frac{J}{\Omega_s^{m_1}}, \nn\\
e^{\beta(\varepsilon_{m_1}-\varepsilon_s)}\rho_{m_1} & = &
e^{\beta(\varepsilon_{m_2}-\varepsilon_s)}\rho_{m_2}
+\frac{Je^{\beta(\varepsilon_{m_1}-\varepsilon_s)}}{\Omega_{m_1}^{m_2}},
\nn\\
e^{\beta(\varepsilon_{m_2}-\varepsilon_s)}\rho_{m_2} & = &
e^{\beta(\varepsilon_{m_3}-\varepsilon_s)}\rho_{m_3}
+\frac{Je^{\beta(\varepsilon_{m_2}-\varepsilon_s)}}{\Omega_{m_2}^{m_3}}, \nn\\
& \vdots &
\nn\\
e^{\beta(\varepsilon_{m_p}-\varepsilon_{-s})}\rho_{m_p} & = &
e^{\beta(\varepsilon_{-s}-\varepsilon_s)}\rho_{-s}
+\frac{Je^{\beta(\varepsilon_{m_p}-\varepsilon_s)}}{\Omega_{m_p}^{-s}},
\nn\\
\hline
\lefteqn{\rho_s=e^{\beta(\varepsilon_{-s}-\varepsilon_s)}\rho_{-s}
+J\sum_{m_i}
\frac{e^{\beta(\varepsilon_{m_i}-\varepsilon_s)}}{\Omega_{m_i}^{m_{i+1}}}}.
\hspace{2cm}
\label{ps}
\eea
In the special case of $H_z=0$, i.e., $\varepsilon_s-\varepsilon_{-s}=0$, Eq.~(\ref{ps}) agrees with previous results.\cite{VILLAIN}

Taking the time derivative of Eq.~(\ref{ps}) and
using $\dot{\rho}_{-s}/\dot{\rho}_s=-1$ we find
\bea
\dot{\rho}_{\pm s}=\pm\frac{\dot{J}}{1+e^{\beta(\varepsilon_{-s}-\varepsilon_s)}}
\sum_{m_i}
\frac{e^{\beta(\varepsilon_{m_i}-\varepsilon_s)}}{\Omega_{m_i}^{m_{i+1}}},
\eea
and thus
\bea
\dot{\rho}_s-\dot{\rho}_{-s} & = &
2\frac{\dot{J}}{1+e^{\beta(\varepsilon_{-s}-\varepsilon_s)}}
\sum_{m_i}
\frac{e^{\beta(\varepsilon_{m_i}-\varepsilon_s)}}{\Omega_{m_i}^{m_{i+1}}}
\nn\\
& = & -2J.
\eea
The solution of the last differential equation is
\be
J(t)=J_0e^{-t/\tau^*},
\ee
with the relaxation time
\be
\tau^*=\frac{1}{1+e^{\beta(\varepsilon_{-s}-\varepsilon_s)}}
\sum_{m_i}\frac{e^{\beta(\varepsilon_{m_i}-\varepsilon_s)}}{\Omega_{m_i}^{m_{i+1}}},\;
H_z\ge 0.
\label{relaxation}
\ee
Finally, the summation $\sum_{\{m_i\}_n}$ in Eq.~(\ref{analytic}) is defined as the summation $\sum_{m_i}$ in Eq.~(\ref{relaxation}) taken only from $m_{\rm initial}$ to $m_{\rm final}+1$, where $\left|m_{\rm initial}\right>,\left|m_{\rm final}\right>$ denote any two neighboring vertices (where paths intersect) in Figs.~\ref{diagram1}, \ref{diagram2}, \ref{diagram3}, and \ref{diagram4} (see below).

\begin{figure}
\begin{center}
    \leavevmode
\epsfxsize=8.5cm
\epsffile{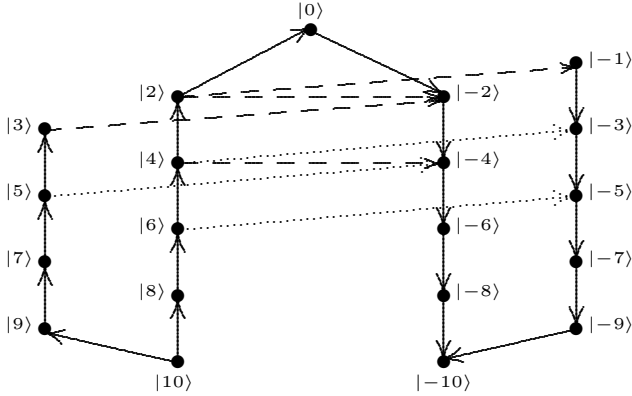}
\end{center}
\caption{Spin relaxation paths (from $m=10$ to $m=-10$) for $0\leq g\mu_B H_z\leq A+13B$. 
Full lines: thermal transitions due to phonons. 
Dashed lines: dominant tunneling transitions due to $B_4$ and $H_x$ terms. Dotted lines: tunneling transitions that lead to satellite peaks [included in the numerical diagonalization of the master equation (\protect\ref{veceq})]. The states where paths intersect are denoted as vertices.}
\label{diagram1}
\end{figure}

\begin{figure}
\begin{center}
    \leavevmode
\epsfxsize=5cm
\epsffile{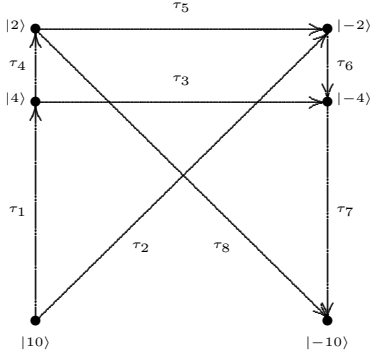}
\end{center}
\caption{Serially reduced diagram associated with Fig.~\protect\ref{diagram1}. In order to understand the analytical evaluation of the relaxation diagram in Fig.~\protect\ref{diagram1} better, tunneling transitions that lead to satellite peaks are excluded.
The relaxation times $\tau_n$ are given in Eq.~(\protect\ref{analytic}). 
For $\left|H_z\right|\protect\lesssim 0.05$ T only the path $\tau_1\rightarrow \tau_3\rightarrow\tau_7$ is dominant (see Figs.~\protect\ref{singlepeak} and \protect\ref{fourpeaks}).}
\label{diag1ser}
\end{figure}

Similarly one can solve the rate equations (\ref{stationary_limit}) for $J<0\Leftrightarrow H_z\le 0$. Then we obtain
\be
\tau^*=\frac{1}{1+e^{\beta(\varepsilon_s-\varepsilon_{-s})}}
\sum_{m_i}\frac{e^{\beta(\varepsilon_{m_i}-\varepsilon_{-s})}}{\Omega_{m_{i+1}}^{m_i}},\;
H_z\le 0,
\ee
which for $H_z=0$ (i.e., $\varepsilon_{-s}=\varepsilon_s$) and steps $\Delta m=\pm 1$ reduces to the result found in Ref.~\onlinecite{VILLAIN}. 

If there is more than one path contributing to the relaxation (which is typically the case in the region between two resonances), we have to account for intersections at vertices. For this we associate with each path a probability current $J_n=\dot{\rho}_n$, and interpret  Eq.~(\ref{analytic}) in terms of a serial circuit with the summands playing the role of ``resistances." This  allows us then to set up  flow diagrams for $J_n$ (see Figs.~\ref{diagram1}, \ref{diag1ser}, and \ref{diagram2}--\ref{diag4ser}), which obey the analog of Kirchhoff's rules:

(K1) $\sum_n J_n=0$: The sum over all incoming and outgoing currents vanishes at a vertex (current conservation).

(K2) $\sum_n J_n\tau_n=\Delta N$: The sum over all voltage drops ($J_n\tau_n$) is equal to the source-drain voltage  $\Delta N=\rho_s-\rho_{-s}$ for any closed path (probability conservation).

The total probability current is given by $J={\Delta {\dot N}}$. 
For every interval 
\be
I_{n_1,n_2}=\left[H_z^{m_{{\rm T},1}m'_{{\rm T},1}},H_z^{m_{{\rm T},2}m'_{{\rm T},2}}\right]
\label{interval}
\ee
[see Eq.~(\ref{condition})], where $n_1=m_{{\rm T},1}+m'_{{\rm T},1}$, $n_2=m_{{\rm T},2}+m'_{{\rm T},2}$, $0\le n_1=n_2+1\le 3$, a set of equations is given by the rules (K1) and (K2). For every set we derive the relaxation time $\tau_{n_1,n_2}^*=\Delta N/J$.

Figure \ref{diagram1} shows the complete, Fig.~\ref{diag1ser} its serially reduced flow diagram for $ 0\le H_z\le \frac{1}{g\mu_B}\left(A+13B\right)$. From (K1) we get
\bea
J=J_1+J_2&,&\quad J_2+J_5=J_6 \, ,\nn\\
J_1=J_3+J_4&,&\quad J_3+J_6=J_7
\, ,\nn\\
J_4=J_5+J_8&,&\quad J_7+J_8=J \, ,\nn
\eea
while from (K2) we get
\bea
\Delta N & = & J_1\tau_1+J_3\tau_3+J_7\tau_7 \, ,
\nn\\
J_3\tau_3 & = & J_4\tau_4+J_5\tau_5+J_6\tau_6 \, , \nn\\
J_2\tau_2 &
= & J_1\tau_1+J_4\tau_4+J_5\tau_5 \, , \nn\\
J_8\tau_8 & = &
J_5\tau_5+J_6\tau_6+J_7\tau_7 \, . \nn
\eea
From these equations we obtain

$
\tau_{0,1}^*(H_z)=(\tau_4\tau_1\tau_5\tau_2+\tau_8\tau_4\tau_1\tau_2+
\tau_8\tau_1\tau_5\tau_2+\tau_8\tau_4\tau_1\tau_6+\tau_4\tau_7\tau_5\tau_2+
\tau_4\tau_7\tau_2\tau_6+\tau_8\tau_7\tau_5\tau_6+\tau_8\tau_7\tau_2\tau_4+
\tau_4\tau_7\tau_5\tau_6+\tau_8\tau_7\tau_2\tau_3+\tau_4\tau_7\tau_2\tau_3+
\tau_8\tau_4\tau_3\tau_7+\tau_8\tau_3\tau_5\tau_7+\tau_4\tau_3\tau_5\tau_7+
\tau_8\tau_3\tau_2\tau_6+\tau_8\tau_4\tau_3\tau_2+\tau_4\tau_3\tau_2\tau_6+
\tau_8\tau_4\tau_3\tau_6+\tau_8\tau_3\tau_5\tau_2+\tau_8\tau_3\tau_5\tau_6+
\tau_8\tau_7\tau_5\tau_2+\tau_8\tau_1\tau_2\tau_6+\tau_8\tau_1\tau_5\tau_6+
\tau_3\tau_1\tau_5\tau_6+\tau_8\tau_3\tau_1\tau_6+\tau_7\tau_1\tau_5\tau_2+
\tau_3\tau_1\tau_2\tau_6+\tau_3\tau_1\tau_5\tau_7+\tau_7\tau_1\tau_2\tau_6+
\tau_7\tau_1\tau_5\tau_6+\tau_7\tau_1\tau_2\tau_4+\tau_8\tau_6\tau_1\tau_7+
\tau_4\tau_1\tau_5\tau_7+\tau_8\tau_4\tau_1\tau_7+\tau_8\tau_3\tau_1\tau_2+
\tau_7\tau_1\tau_2\tau_3+\tau_8\tau_3\tau_1\tau_7+\tau_3\tau_1\tau_5\tau_2+
\tau_8\tau_1\tau_5\tau_7+\tau_8\tau_7\tau_2\tau_6+\tau_8\tau_4\tau_7\tau_6+
\tau_4\tau_3\tau_5\tau_6+\tau_4\tau_3\tau_5\tau_2+\tau_4\tau_1\tau_2\tau_6+
\tau_4\tau_1\tau_5\tau_6)/(\tau_8\tau_5\tau_2+\tau_8\tau_5\tau_6+
\tau_8\tau_3\tau_2+\tau_8\tau_4\tau_3+\tau_4\tau_5\tau_6+
\tau_8\tau_4\tau_6+\tau_8\tau_4\tau_2+\tau_4\tau_2\tau_6+
\tau_4\tau_7\tau_6+\tau_8\tau_3\tau_5+\tau_7\tau_2\tau_6+
\tau_7\tau_5\tau_6+\tau_7\tau_5\tau_2+\tau_7\tau_2\tau_3+
\tau_4\tau_1\tau_5+\tau_3\tau_1\tau_5+\tau_8\tau_1\tau_5+\tau_8\tau_3\tau_1+
\tau_3\tau_2\tau_6+\tau_3\tau_5\tau_6+\tau_3\tau_5\tau_2+\tau_3\tau_5\tau_7+
\tau_4\tau_3\tau_7+\tau_4\tau_3\tau_5+\tau_8\tau_6\tau_1+\tau_3\tau_1\tau_7+
\tau_7\tau_1\tau_6+\tau_4\tau_1\tau_6+\tau_3\tau_1\tau_6+\tau_4\tau_1\tau_7+
\tau_8\tau_4\tau_1+\tau_8\tau_2\tau_6+\tau_4\tau_5\tau_2+\tau_4\tau_3\tau_6+
\tau_2\tau_4\tau_7+\tau_7\tau_1\tau_5).
$

When $\tau_{0,1}^*$ is plotted as function of $H_z$ there is no visible difference between the exact $\tau$ obtained in Sec.~\ref{relax} and this approximate $\tau^*$, which confirms that the diagram in Fig.~\ref{diagram1} contains the physically relevant relaxation paths for the interval $I_{0,1}$. Similar results are obtained for the other intervals, whose diagrams and calculations are shown in Appendix \ref{Kirchhoff}. 

Finally, near a resonance [$\left|\delta H_z\right|<w'$, see Eq.~(\ref{width})] the above expression for $\tau^*$ [Eq.~(\ref{analytic})] strongly simplifies since we find that there is only one dominant relaxation path which involves only one tunneling channel. This finally explains why the peak shape is given by a single Lorentzian. We call the five strongest broadened resonances in 
Figs.~\ref{overall}--\ref{overall3} the main resonances. For every main resonance $n$ we have identified [using Eq.~(\ref{analytic})] its dominant path and its associated tunneling channel 
between the states $\left|m_{\rm T}\right>$ and $\left|m_{\rm T}'\right>$. These states are
{\setlength{\arraycolsep}{1cm}
\be
\begin{array}{ccc}n & m_{\rm T} & m'_{\rm T} \\\hline
0 & 4 & -4 \\
1 & 3 & -2 \\
2 & 5 & -3 \\
3 & 4 & -1 \\
4 & 6 & -2
\end{array}
\label{dominant_channel}
\ee
}

Our calculation of the intermediate relaxation times $\tau_n$ provides a further prediction which could be tested with NMR techniques of the type described in Ref.~\onlinecite{Lascialfari}.

\subsection{Satellite peaks}

Beside the main resonances there are also other narrower resonances (see Figs.~\ref{overall}-\ref{overall3}) that are a direct consequence of the fourth-order anisotropy constant $B$ [see Eq.~(\ref{H_a})]. Indeed, if the plots around one peak are magnified further, satellite peaks become visible (see Figs.~\ref{sat2deg2}--\ref{sat4deg3}).
In order to understand the occurrence of these satellite peaks it is instructive to look at Fig.~\ref{diagram3} below. There are several paths which can be used in the relaxation process. As we include the fourth-order anisotropy term, $-BS_z^4$, the resonance condition is not the same for every level [see Eq.~(\ref{condition})]. Hence, very narrow peaks show up, which can be seen only at high resolution. In Fig.~\ref{diagram3} several additional tunneling paths, some of which are responsible for the satellite peaks in Figs.~\ref{sat2deg2} and \ref{sat2deg3}, have to be drawn (represented by the dotted lines in Fig.~\ref{diagram3}). For example, the tunnel splitting energy of the path from $\left|4\right>$ to $\left|-2\right>$ is proportional to $H_xB_4H_x$ (third-order perturbation), where the ordering of the factors corresponds to the chosen path. Due to the presence of $H_x^2$ the width of the satellite peak (see next section) depends on the misalignment angle $\theta$. If one takes a close look at our high resolution plots this difference between Fig.~\ref{sat2deg2} and Fig.~\ref{sat2deg3} is observable. It must be noted that we consider only tunnel splitting energies up to second order in $B_4$ and third order in $H_x$ (also combinations such as $B_4^2H_x^3$) for all the main and satellite peaks. Narrower satellite peaks are neglected.\cite{Teemu} The distance $d_{m_1m_1{'}}^{m_2m_2{'}}$ between a satellite peak and its associated main peak caused by a main resonance is given by Eq.~(\ref{condition}),
\bea
d_{m_1m_1{'}}^{m_2m_2{'}} & = & \left|H_z^{m_1m_1{'}}-H_z^{m_2m_2{'}}\right| \nn\\
& = & \left|\frac{nB}{g\mu_B}\left(m_1^2+m_1{'}^2-m_2^2-m_2{'}^2\right)\right|,
\eea
where $m_1$, $m_1{'}$ ($m_2$, $m_2{'}$) are responsible for the satellite (main) peak, and $n=m_1+m_1{'}=m_2+m_2{'}$. It would be interesting to search experimentally for these satellite peaks, which requires a higher experimental resolution of the peaks than achieved so far.

\begin{figure}
  \begin{center}
    \leavevmode
\epsfxsize=8.5cm
\epsffile{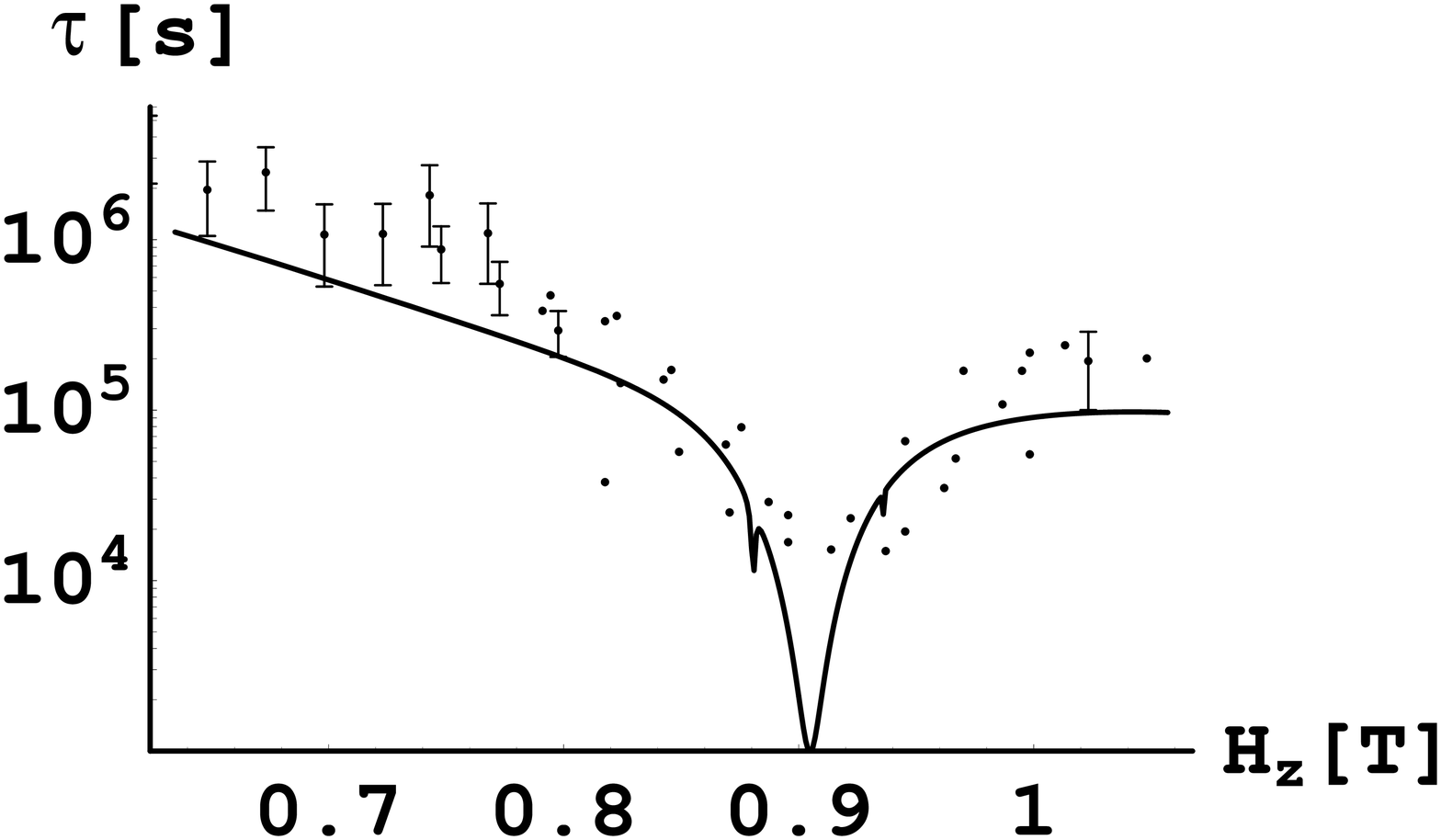}
\end{center}
\caption{Full line: semilogarithmic plot of calculated relaxation time $\tau$ as function of  magnetic field $H_z$ at $T=1.9$ K in the interval $3A/2g\mu_B\le H_z\le 5A/2g\mu_B$ with a higher resolution. The tunneling transition from $\left|5\right>$ and $\left|-3\right>$ is responsible for the main peak. Two satellite peaks are visible. The left (right) one is due to the tunneling channel between $\left|4\right>$ and $\left|-2\right>$ ($\left|6\right>$ and $\left|-4\right>$). Here $\theta=2^\circ$ has been chosen. Dots and error bars: data taken from Ref.~\protect\onlinecite{Thomas}.}
\label{sat2deg2}
\end{figure}

\begin{figure}
  \begin{center}
    \leavevmode
\epsfxsize=8.5cm
\epsffile{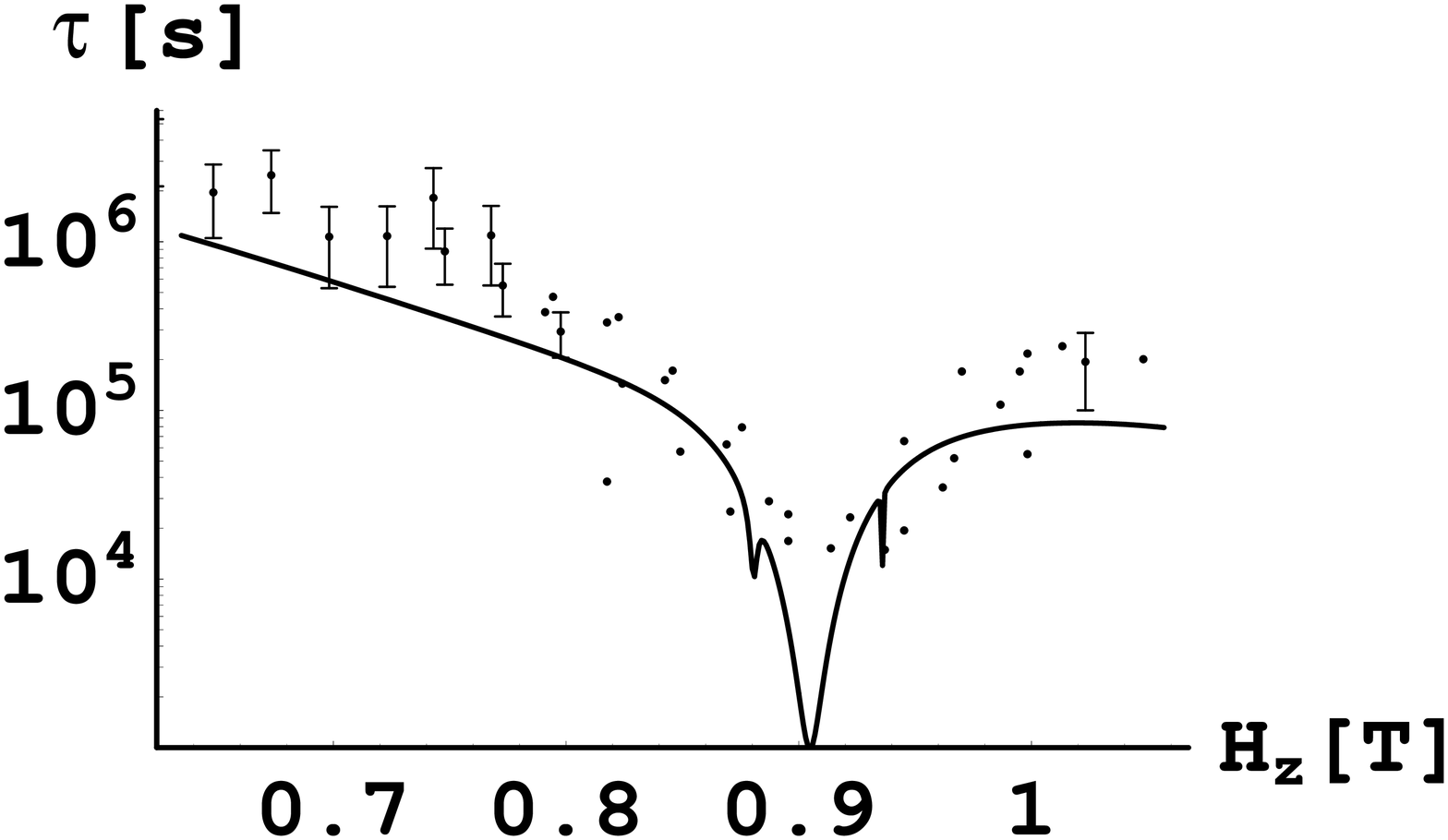}
\end{center}
\caption{Same plot as in Fig.~\protect\ref{sat2deg2}, but with a misalignment angle of $\theta=3^\circ$. Dots and error bars: data taken from Ref.~\protect\onlinecite{Thomas}.}
\label{sat2deg3}
\end{figure}

\begin{figure}
  \begin{center}
    \leavevmode
\epsfxsize=8.5cm
\epsffile{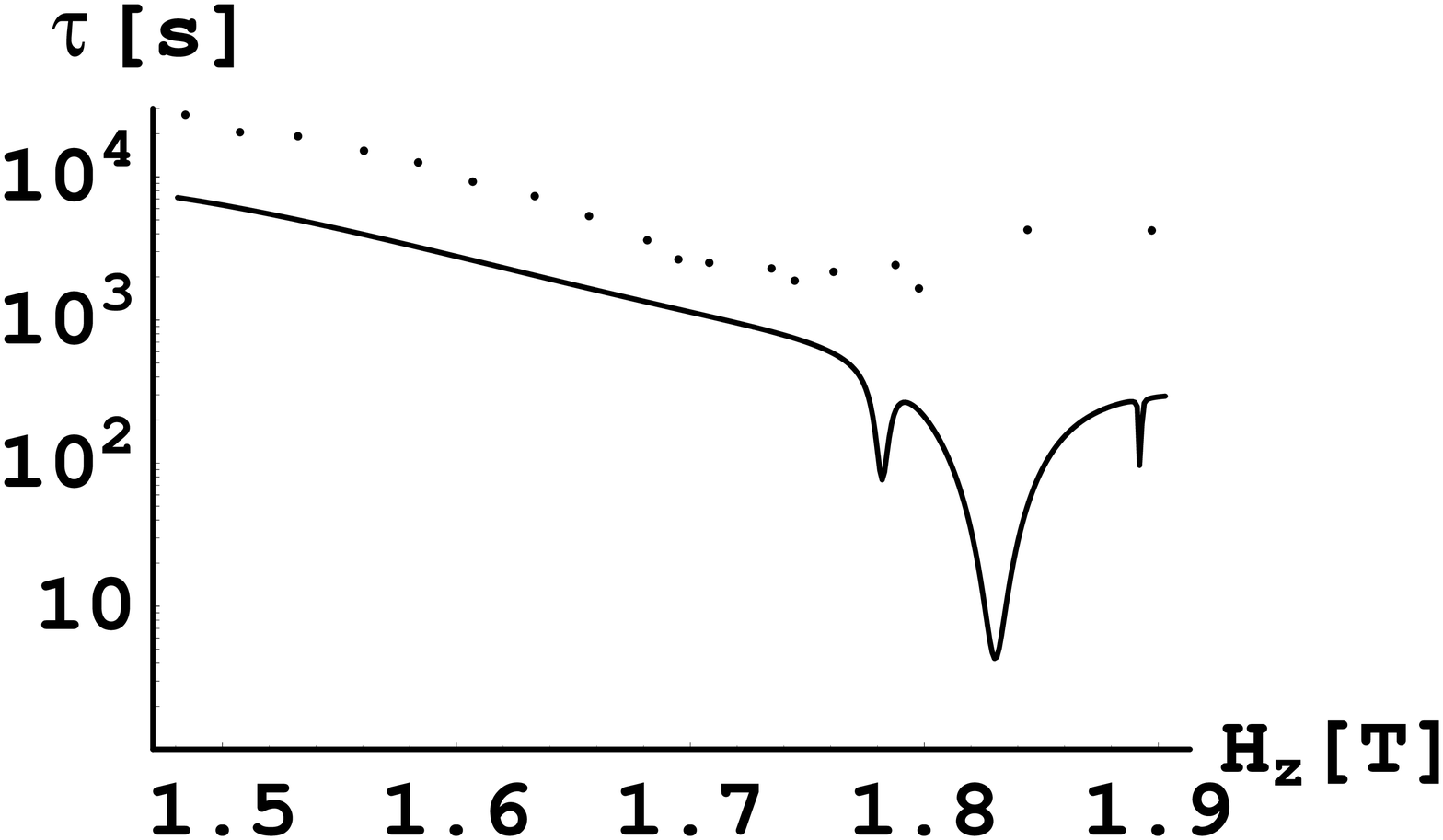}
\end{center}
\caption{Full line: semilogarithmic plot of calculated relaxation time $\tau$ as function of  magnetic field $H_z$ at $T=1.9$ K in the interval $7A/2g\mu_B\le H_z\le 9A/2g\mu_B$ with a higher resolution. The tunneling transition from $\left|6\right>$ and $\left|-2\right>$ is responsible for the main peak. Two satellite peaks are visible. The left (right) one is due to the tunneling channel between $\left|5\right>$ and $\left|-1\right>$ ($\left|7\right>$ and $\left|-3\right>$). Here $\theta=2^\circ$ has been chosen. Dots and error bars: data taken from Ref.~\protect\onlinecite{Thomas}.}
\label{sat4deg2}
\end{figure}

\begin{figure}
  \begin{center}
    \leavevmode
\epsfxsize=8.5cm
\epsffile{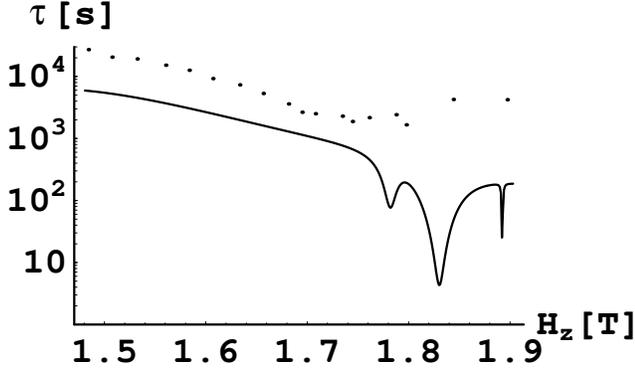}
\end{center}
\caption{Same plot as in Fig.~\protect\ref{sat4deg2}, but with a misalignment angle of $\theta=3^\circ$. Dots and error bars: data taken from Ref.~\protect\onlinecite{Thomas}.}
\label{sat4deg3}
\end{figure}


\section{Width of the Lorentzians}
\label{lorentzian}

In this section we give a physical interpretation of the effective half-width of the Lorentzian peaks in our plots. In order to get an expression for the width of our main and satellite peaks consider a Lorentzian $\Gamma(H_z)$ with linewidth $w$ (see Fig.~\ref{truncation}). If the upper part of this Lorentzian is cut off (where the curve is already very narrow) and both ends are connected by a horizontal line one obtains a curve that still has  the same single Lorentzian shape for all practical purposes but now with an effective linewidth $w'>w$. Changing the tunnel matrix element $E_{mm'}$ results in a different truncation of the Lorentzian, thus changing the effective linewidth $w'$. We shall now estimate the effective linewidth $w'$ and compare it with the one obtained from the exact $1/\tau$. Taking only the largest terms of Eq.~(\ref{analytic}) gives a rough approximation of the relaxation time near a resonance where the states $\left|m\right>$ and $\left|m'\right>$ are degenerate,
\bea
\tau' & = &
\frac{1}{1+e^{\beta(\varepsilon_{-s}-\varepsilon_s)}}
\left(\frac{e^{\beta(\varepsilon_{m+2}-\varepsilon_s)}}{W_{m,m+2}}
+\frac{e^{\beta(\varepsilon_{m'}-\varepsilon_s)}}{W_{m'-2,m'}}\right.
\nn\\ 
& & +\left.\frac{e^{\beta(\varepsilon_{m}-\varepsilon_{-s})}}
{\Gamma_{m}^{m'}}\right),
\eea
Using the detailed balance relation
\be
\frac{W_{m,m+2}}{W_{m+2,m}}=e^{\beta(\varepsilon_{m+2}-\varepsilon_m)}
\ee
we obtain the following approximation:
\be
\tau'=\frac{e^{\beta(\varepsilon_{m+2}-\varepsilon_s)}}{1+ e^{\beta(\varepsilon_{s}-\varepsilon_s)}}
\left(\frac{2}{W_{m,m+2}}+\frac{1}{\Gamma_{m}^{m'}}\right),
\ee
where we assumed that $W_{m,m+2}\approx W_{m'-2,m'}$.\cite{detailed_balance} In the limit $\xi_{mm'}\rightarrow 0$ the phonon-damped tunneling rate $\Gamma_{m}^{m'}$ is much larger than $W_{m,m+2}$, so
\be
\lim_{\xi_{mm'}\rightarrow 0}\tau'\approx
\frac{2e^{\beta(\varepsilon_{m+2}-\varepsilon_s)}}{\left(1+e^{\beta(\varepsilon_{-s}-\varepsilon_s)}\right)W_{m,m+2}}.
\ee
The half-width of $\tau'(H_z)$, denoted by $w'$,  is then determined by the condition $\tau'(w'/2)=\tau'(0)/2$. This condition is fulfilled when
\be
\Gamma_{m}^{m'}=\frac{W_{m,m+2}}{2}.
\ee
Thus we obtain the expression for the effective linewidth $w'$,
\be
w'=\frac{2\sqrt{W_{m}+W_{m'}}}{|m-m'|g\mu_B}\left[\frac{E_{mm'}^2}
{W_{m,m+2}}-\frac{\hbar^2\left(W_{m}+W_{m'}\right)}{4}\right]^{1/2}.
\ee
Since the height $E_{mm'}^2/\hbar^2\left(W_{m}+W_{m'}\right)$ of the Lorentzian $\Gamma_{m}^{m'}$ is very large compared to its linewidth $|m-m'|g\mu_Bw/\hbar=\left(W_{m}+W_{m'}\right)/2$ and $W_{m}+W_{m'}\approx 2W_{m,m+2}$ for the dominant paths (see Sec.~\ref{paths}) we get the following reasonably accurate approximation for the effective linewidths in our plots:
\be
w'=\frac{2^{3/2}E_{mm'}}{|m-m'|g\mu_B}.
\label{width}
\ee
Comparison with our exact calculations of the relaxation time shows that $w'$ of Eq.~(\ref{width}) gives a very good estimate for the effective linewidth of the peaks in our plots (see Figs.~\ref{overall}--\ref{fourpeaks} and \ref{sat2deg2}--\ref{truncation}).

\begin{figure}[htb]
  \begin{center}

\leavevmode
\epsfxsize=8.5cm
\epsffile{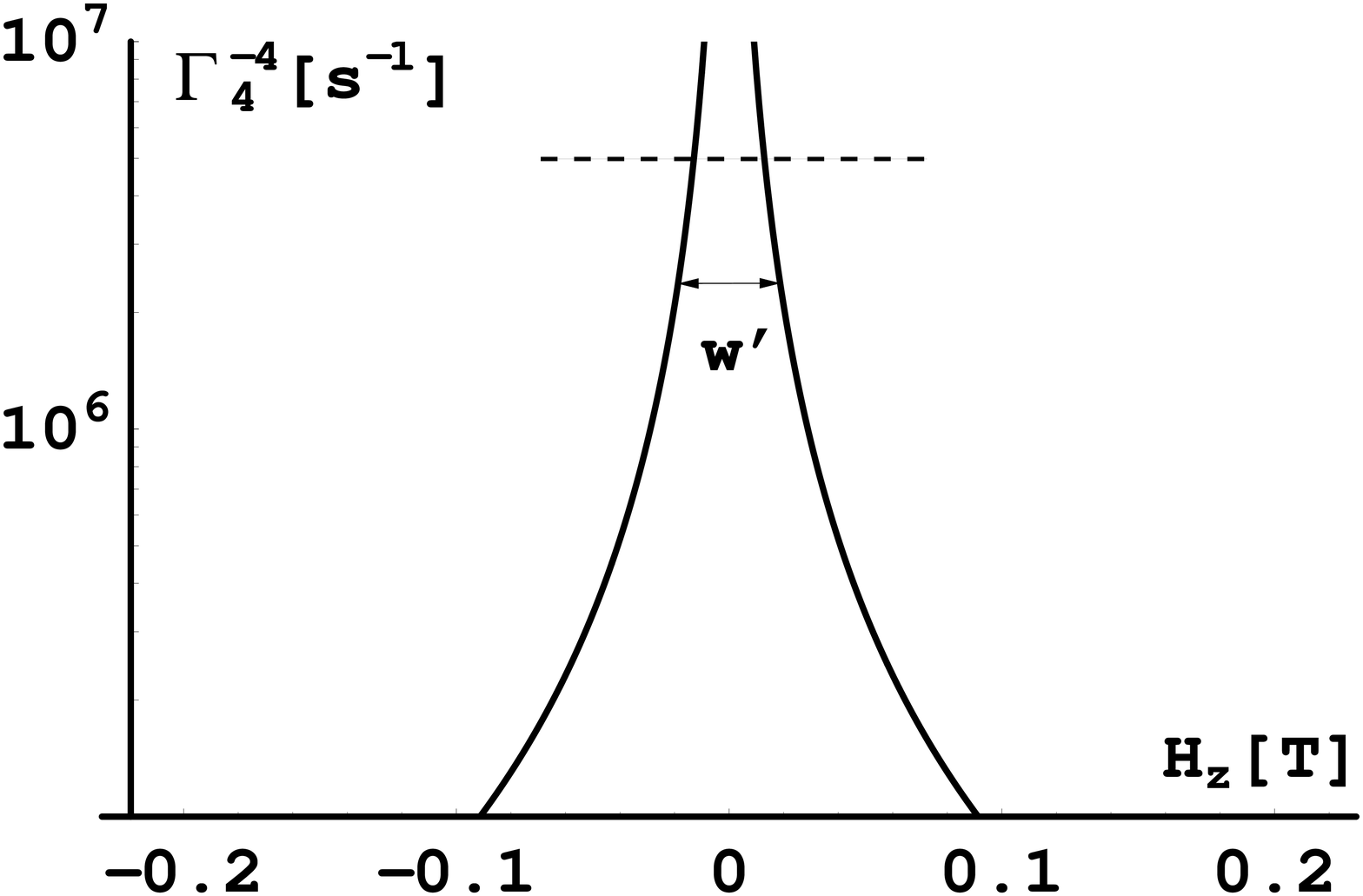}

\end{center}
\caption{Truncated Lorentzian $\Gamma_4^{-4}$ with
$A/k_B=0.56$ K, $B/k_B=1.3\times 10^{-3}$ K, and $B_4/k_B=14.4\times 10^{-5}$ K,
$\theta=1^\circ$, and  $c=2.0\times 10^3$ m/s, $w'=37.4$ mT ($w'$ agrees very well with the width of the Lorentzian in Fig.~\protect\ref{singlepeak}), and
$\Gamma_4^{-4}(w'/2)=2.4\times 10^6$ s$^{-1}$. The truncation is indicated by the dashed line.}
\label{truncation}
\end{figure}


\section{Conclusion}

We have presented a comprehensive theoretical description of spin relaxation due to phonon-induced transitions and tunnel resonances. Deriving a generalized master equation (in Born and Markoff approximation) we obtain an exact numerical evaluation of the overall relaxation time $\tau$ as function of the longitudinal magnetic field $H_z$ comprising Lorentzian-shaped peaks. In order to perform this evaluation we calculate the phonon-assisted transition rates of the spins, the spin-phonon coupling constants, and the tunnel splitting energy, for which a generalized formula is derived. The fourth-order diagonal terms in the Hamiltonian give rise to satellite peaks, the experimental observation of which requires a higher resolution of $\tau(H_z)$ than achieved so far.
Our approximate analytical solution of the master equation yields a clear physical understanding of the relaxation process by revealing the relaxation paths that are followed by the spin. This solution provides the prediction of all involved intermediate relaxation times $\tau_n$, which can be tested experimentally.
The results of our model calculation agree well with {\it all} known data. For the first time we have been able to get agreement between theory and the entire relaxation curve. In addition, we have obtained reasonable agreement between theory and four single resonance peaks recently measured to high accuracy at four different temperatures. The formalism presented in this work has been applied to the specific parameter values of Mn$_{12}$, but many results derived here are generally valid and can be used for similar spin systems as well.


\acknowledgments{
We are grateful to H.~Schoeller and T.~Pohjola for useful comments.
This work has been supported  by the Swiss National Science Foundation.}


\begin{appendix}

\section{Spin-phonon rates}
\label{sp}

In order to evaluate the spin-phonon rates $W_{mn}$ of Eq.~(\ref{FGR}) we first change to the Fourier representation. If ${\bf q}$ is a phonon wave vector, we can write {\bf u}({\bf x}) as follows:
\be
{\bf u}({\bf
x})=\frac{1}{\sqrt{N}}\sum\limits_{\bf q}{\bf u}({\bf
q})e^{ i{\bf q\cdot
x}},
\ee
with $N$ being the number of unit cells.
Hence
\be
\epsilon({\bf x})=\frac{ i}{\sqrt{N}} \sum\limits_{\bf q}
\left[
\begin{array}{ccc}
 q_xu_x({\bf q}) & q_xu_y({\bf q}) & q_xu_z({\bf q}) \\

q_yu_x({\bf q}) & q_yu_y({\bf q}) & q_yu_z({\bf q}) \\
 q_zu_x({\bf q}) &
q_zu_y({\bf q}) & q_zu_z({\bf q})
 \end{array} \right] e^{ i{\bf q\cdot
x}}.
\ee
After (anti)symmetrization, these matrix elements can be inserted into the expression (\ref{Spin-phonon}),
\bea
{\cal H}_{\rm sp} & = &
\frac{1}{\sqrt{N}}
\sum\limits_{j,\bf
q} i\left\{
\frac{1}{2}g_1[q_xu_x({\bf q})-q_yu_y({\bf q})]\otimes
(S_+^2+S_-^2)
\right.\nn\\
& & +\left.\frac{ i}{8}g_2[q_xu_y({\bf
q})+q_yu_x({\bf q})]\otimes
(S_-^2-S_+^2) \right.\nn\\
& &
+\left.\frac{1}{8}g_3[q_xu_z({\bf q})+q_zu_x({\bf q}) \right.\nn\\
& &
-\left. i(q_yu_z({\bf q})+q_zu_y({\bf q}))]\otimes \{S_+,S_z\}
\right.\nn\\
& & +\left.\frac{1}{8}g_3[q_xu_z({\bf q})+q_zu_x({\bf q})
\right.\nn\\
& & +\left. i(q_yu_z({\bf q})+q_zu_y({\bf q}))]\otimes
\{S_-,S_z\} \right.\nn\\
& & +\left.\frac{1}{8}g_4[q_xu_z({\bf
q})+q_zu_x({\bf q}) \right.\nn\\
& & -\left. i(q_yu_z({\bf q})+q_zu_y({\bf
q}))]\otimes \{S_+,S_z\} \right.\nn\\
& & +\left.\frac{1}{8}g_4[q_xu_z({\bf
q})-q_zu_x({\bf q}) \right.\nn\\
& & +\left. i(q_yu_z({\bf q})-q_zu_y({\bf
q}))]\otimes \{S_-,S_z\}
\right\}e^{ i{\bf q\cdot
R}_j}.
\label{sp-ph}
\eea
${\bf R}_j$ are the positions of the Mn$_{12}$ molecules.

We proceed with the canonical transformation $({\bf u},{\bf p})\to ({\bf c}^\dagger,{\bf c})$. ${\bf c}^{(\dagger)}={\bepsilon}_{\bf q}c_{\bf q}^{(\dagger)}$  annihilates (creates) a phonon with wave vector ${\bf q}$ and polarization ${\bepsilon}_{\bf q}$, and 
\be 
{\bf u}({\bf q})=\sqrt{\frac{\hbar}{2M\omega_{{\bf q}}}}\left({\bf c}^\dagger+{\bf c}\right), \label{canonical} 
\ee 
where $M$ is the mass per unit cell.
Inserting Eq.~(\ref{canonical}) into Eq.~(\ref{sp-ph}) and considering only the spin of the  Mn$_{12}$ molecule at ${\bf R}_j=0$ yields
\bea
{\cal H}_{\rm sp} & =
&
\sum\limits_{{\bf q}} i\sqrt{\frac{\hbar}{2NM\omega_{{\bf q}}}} \nn\\
& &
\times\left\{
\frac{1}{2}g_1[q_x(c_x^\dagger+c_x)-q_y(c_y^\dagger+c_y)]\otimes
(
S_+^2+S_-^2) \right.\nn\\
& &
+\left.\frac{ i}{8}g_2[q_x(c_y^\dagger+c_y)+q_y(c_x^\dagger+c_x)]\otimes
(S_-^2-
S_+^2) \right.
\nn\\
& & +\left.\frac{1}{8}g_3[(q_x- i
q_y)(c_z^\dagger+c_z) \right.\nn\\
& & +\left.q_z(c_x^\dagger+c_x - i
c_y^\dagger- i c_y)]\otimes \{S_+,S_z\}
\right. \nn\\
& &
+\left.\frac{1}{8}g_3[(q_x+ i q_y)(c_z^\dagger+c_z) \right.\nn\\
& &
+\left.q_z(c_x^\dagger+c_x+ i c_y^\dagger+ i c_y)]\otimes
\{S_-,S_z\}
\right. \nn\\
& & +\left.\frac{1}{8}g_4[(q_x- i
q_y)(c_z^\dagger+c_z) \right.\nn\\
& & -\left.q_z(c_x^\dagger+c_x- i
c_y^\dagger- i c_y)]\otimes \{S_+,S_z\}
\right.\nn\\
& &
+\left.\frac{1}{8}g_4[(q_x+ i q_y)(c_z^\dagger+c_z) \right.\nn\\
& &
-\left.q_z(c_x^\dagger+c_x+ i c_y^\dagger+ i c_y)]\otimes
\{S_-,S_z\}
\right\}.
\eea
This expression can be used to evaluate the transition probability. We employ the following standard relations:
\bea
c\left|n\right> & = & \sqrt{n}\left|n-1\right>,
\nn\\
c^\dagger\left|n\right> & = & \sqrt{n+1}\left|n+1\right>,
\nn\\
S_-\left|s,m\right> & = & \sqrt{(s+m)(s-m+1)}\left|s,m-1\right>,
\nn\\
S_+\left|s,m\right> & = &
\sqrt{(s-m)(s+m+1)}\left|s,m+1\right>.
\eea
The transition rate $W_{-2}=W_{m-2,m}$ [see Eq.~(\ref{FGR})] for $m\to m-2$ ($\varepsilon_{m-2}\gtrless\varepsilon_m$) can now be calculated in second quantization ($n_\alpha=n_{{\bf q},\alpha}$, $\alpha=x,y,z$ denotes the number of phonons with wave vector ${\bf q}$, polarization mode $\lambda$, and oscillation direction $\alpha$, and the thermal average over phonons is left implicit),
\bea
W_{-2} & = &
\frac{2\pi}{\hbar}\sum_{{\bf q}'}
|\left<n_{{\bf q}'}\mp 1,m-2\left|\cH_{\rm sp}\right|n_{{\bf q}'},m\right>|^2\delta'_\pm \nn\\
& = & \sum_{{\bf q}}\frac{\pi}{NM\omega_{{\bf q}}}
\nn\\
& &
\left[\frac{g_1^2}{4}\left(q_x\left<n_x\mp 1|c_x^{(\dagger)}|n_x\right>-q_y\left
<n_y\mp 1|c_y^{(\dagger)}|n_y\right>\right)^2 \right.\nn\\
& &
\times\left.\left|\left<m-2|S_-^2|m\right>\right|^2 \right.
\nn\\
& &
+\left.\frac{g_2^2}{64}\left(q_x\left<n_y\mp 1|c_y^{(\dagger)}|n_y\right>
+q_y\left<n_x\mp 1|c_x^{(\dagger)}|n_x\right>\right)^2 \right.\nn\\
& &
\times\left.\left|\left<m-2|S_-^2|m\right>\right|^2\right] \delta_\pm
\nn\\
& = & \frac{1}{4}\sum_{{\bf q}}\frac{\pi s_{-2}}{NM\omega_{{\bf q}}}\left(n_{{\bf q}}+{1\atop 0}\right) \nn\\
& & \times\left[g_1^2(q_x-q_y)^2+\frac{g_2^2}{16}(q_x+q_y)^2\right]\delta_\pm,
\eea
where $s_{-2}=(s+m)(s-m+1)(s+m-1)(s-m+2)$, and $\delta^({'}^)_\pm=\delta(\pm(\varepsilon_{m-2}-\varepsilon_m)-\hbar\omega_{{\bf q }^({'}^)})$.

With the approximation $g_1=A\approx g_2$ and the thermal average 
$\left<n_{{\bf q}}\right>=1/(e^{\beta\hbar\omega_q}-1)$ one obtains
\be
W_{-2}=\frac{1}{4}\sum_{{\bf q}}\frac{\pi
A^2s_{-2}}{NM\omega_q}\;
\frac{(q_x-q_y)^2+\frac{1}{16}(q_x+q_y)^2}{\pm\left(e^{
\pm\beta\hbar\omega_q}-
1\right)}\delta_\pm.
\ee
As a next step the sum is replaced by an integral $\left[((1/N)\sum_{{\bf q}}\to(a^3/(2\pi)^3)\int\;d^3q\right]$ and the density $\rho=M/a^3$ is inserted,
\be
W_{-2}=\frac{A^2s_{-2}}{32\pi^2\rho}\int\frac{d^3q}{\omega_q}\;
\frac{(q_x-q_y)^2+\frac{1}{16}(q_x+q_y)^2}{\pm\left(e^{\pm\beta\hbar\omega_q}
-1\right)}\delta_\pm.
\ee
After changing to spherical coordinates one gets
\be
W_{-2}=\frac{17A^2s_{-2}}{192\pi\rho}\int_0^\infty
\frac{dq}{\omega_q} \;\frac{q^4}{\pm\left(e^{\pm\beta\hbar\omega_q}-1\right)}\delta_\pm.
\ee
Assuming a linear dispersion relation $\omega_q=cq$, where $c$ is the sound velocity, and using $\varepsilon=\hbar\omega_q=\hbar cq$ one obtains
\bea
W_{-2} & = & \frac{17A^2s_{-2}}{192\pi\rho c^5\hbar^4}
\int_0^\infty
d\varepsilon\;\frac{\varepsilon^3}{\pm\left(e^{\pm\beta\hbar\omega_q}-1\right)}\delta_\pm
\nn\\
& = & \frac{17A^2s_{-2}}{192\pi\rho c^5\hbar^4}\;
\frac{(\varepsilon_{m-2}-\varepsilon_m)^3}{e^{\beta(\varepsilon_{m-2}-\varepsilon_m)}-1}.
\eea
In the same way we get
\be
W_{+2}=\frac{17A^2s_{+2}}{192\pi\rho c^5\hbar^4}\;
\frac{(\varepsilon_{m+2}-\varepsilon_m)^3}{e^{\beta(\varepsilon_{m+2}-\varepsilon_m)}-1},
\ee
with $s_{+2}=(s-m)(s+m+1)(s-m-1)(s+m+2)$.

The transition rates for $m\to m\pm 1$ can be calculated in the same manner as above with $g_4=2A\approx g_3$,
\be
W_{\pm 1}=\frac{A^2s_{\pm 1}}{12\pi\rho c^5\hbar^4}\;
\frac{(\varepsilon_{m\pm 1}-\varepsilon_m)^3}
{e^{\beta(\varepsilon_{m\pm 1}-\varepsilon_m)}-1},
\ee
where $s_{\pm 1}=(s\mp m)(s\pm m+1)(2m\pm 1)^2$, and $\rho=1.83\times 10^3$ kg/m$^3$.\cite{Lis}


\section{Level splitting}
\label{resolvent}
In this appendix we derive a formula for the tunnel splitting energy which is applicable to
potentials $V_{m_i,m_{i+1}}\in \Bbb{R}$ with arbitrary $\Delta m=m_i-m_{i+1}$ ($m>m_i>m_{i+1}>m'$, $i=1,\ldots,N-1$). According to Kato's theory \cite{Messiah} the expansion of the resolvent
\be
G(z)=\frac{1}{z-H_0-\lambda V}
\ee
leads to a rigorous treatment of the perturbation theory, which is very useful to evaluate high-order perturbation terms. We use the notation of Messiah.\cite{Messiah}

Let $N$ be the order of the perturbation. Then the projection operator $P=\sum_{m}\left|m\right>\left<m\right|$, consisting of the degenerate states $\left\{\left|m\right>\right\}$, and the operator $\left(H-E_a^0\right)P$ are expanded as follows:
\[
P=P_0+\sum\limits_{N=1}^\infty \lambda^N A^{(N)},\;
\left(H-E_a^0\right)P=\sum\limits_{N=1}^\infty \lambda^N B^{(N)},
\]
with
\bea
A^{(N)} & = & -\sum\limits_{(N)}S^{k_1}VS^{k_2}V\cdots VS^{k_{N+1}}, \nn\\
B^{(N)} & = & -\sum\limits_{(N-1)}S^{k_1}VS^{k_2}V\cdots VS^{k_{N+1}},\nn
\eea
where
\bea
S^k & = &\left\{
\begin{array}{rl} -P_0, & \text{if } k=0, \\ \frac{Q_0}{a^k}, & \text{if }
k\ge 1,\end{array}
\right.,
\nn\\
Q_0 & = & 1-P_0,\quad \frac{Q_0}{a^k}=Q_0\frac{1}{\left(E_a^0-H_0\right)^k}Q_0, \nn
\eea
and the sum $\sum_{(N)}$ has to be taken over all combinations $k_1,k_2,\ldots,k_{N+1}$ with the restriction $k_1+k_2+\cdots+k_{N+1}=N$.

The following general secular equation must be solved:
\be
\text{det}\left(H_a-\chi
K_a\right)=\text{det}\left(C_a\right)=0,
\ee
where we have introduced the abbreviation $C_a=H_a-\chi K_a$. The $\chi$ are the eigenvalues of the perturbed states. $H_a$ and $K_a$ are defined by
\bea
H_a & = & P_0HPP_0=E_a^0K_a+P_0\sum\limits_{N=1}^\infty \lambda^N B^{(N)}P_0, \\
K_a & = & P_0PP_0=P_0+P_0\sum\limits_{N=1}^\infty \lambda^N A^{(N)}P_0.
\eea
Thus we have now
\bea
C_a & = & \left(E_a^0-\chi\right)P_0+\left(E_a^0-\chi\right)\sum\limits_{N=1}^\infty \lambda^N P_0A^{(N)}P_0 \nn\\
& & +\left.\sum\limits_{N=1}^\infty \lambda^N P_0B^{(N)}P_0.\right.
\label{pt}
\eea
Equation (\ref{pt}) is the general formula for finding the perturbed eigenvalues and eigenstates. We apply it now to the situation of our two degenerate spin states $\left|m\right>$ and $\left|m'\right>$. The following derivation refers to the off-diagonal elements of Eq.~(\ref{pt}).

The factors $P_0A^{(N)}P_0$ and $P_0B^{(N)}P_0$ do not vanish if $k_1=k_{N+1}=0$. As we look for the lowest-order perturbation that gives a contribution to the tunnel splitting $E_{mm'}$, the projection operators $S^{k_i}$, $i=2,\ldots,N$, must not be equal to $-P_0$, i.e., $k_i\ne 0$, $i=2,\ldots,N$. Hence, we get the following combinations for the lowest-order perturbation,
\bea
\text{for }A^{(N)} & : & \; k_2=k_3=\cdots=k_{i-1}=1,k_i=2, \nn\\
& & \; k_{i+1}=\ldots=k_N=1,\; i=2,\cdots,N, \\
\text{for }B^{(N)} & : & \; k_2=k_3=\cdots=k_N=1.
\eea
In the case of weak perturbation the second term of Eq.~(\ref{pt}) is much smaller than the third one. Thus the secular equation reads as follows:
\bea
C_a & = & \left(E_a^0-\chi\right)P_0+\text{(diagonal elements)} \nn\\
& & +\left.\sum\limits_{N=1}^\infty
\lambda^N \sum\limits_{m_1,\ldots,m_N \atop m_i\ne m,m'}\left|m\right>\frac{V_{m,m_1}}{\varepsilon_m-
\varepsilon_{m_1}} \right.\nn\\
& & \times\left.\prod\limits_{i=1}^{N-1} \frac{V_{m_i,m_{i+1}}}{\varepsilon_m
-\varepsilon_{m_{i+1}}}V_{m_N,m'} \left<m'\right|. \right.
\eea
Thus we arrive at formula (\ref{gsplitting}).


\section{Application of Kirchhoff's rules}
\label{Kirchhoff}

In this appendix we make use of Kirchhoff's rules (K1) and (K2) in order to evaluate the diagrams of the relaxation paths. Each diagram and its evaluation is valid for the interval between two main peaks. The solution $\tau_{n_1,n_2}^*$ of the Kirchhoff equations between the peaks $n_1=m_1+m_1'$ and $n_2=m_2+m_2'$ is not written down explicitly, since it is too lengthy and the calculation is straightforward.

$(1)\;\frac{1}{g\mu_B}\left(A+13B\right)\le
H_z\le
\frac{2}{g\mu_B}\left(A+34B\right)$:
\\
From
$J=J_1+J_2,
J_1=J_3+J_4,
J_4=J_5+J_8,
J_2+J_5=J_6,
J_3+J_6=J_7,
J_7+J_8=J$, and
$\Delta N=J_1\tau_1+J_3\tau_3+J_7\tau_7,
J_3\tau_3=J_4\tau_4+J_5\tau_5+J_6\tau_6,
J_2\tau_2=J_1\tau_1+J_4\tau_4+J_5\tau_5,
J_8\tau_8=J_5\tau_5+J_6\tau_6+J_7\tau_7$,
one can immediately evaluate the relaxation time $\tau_{1,2}^*(H)=\Delta N/J$ (see Figs.~\ref{diagram2} and \ref{diag2ser}).

\begin{figure}[h]
  \begin{center}

\leavevmode
\epsfxsize=8cm
\epsffile{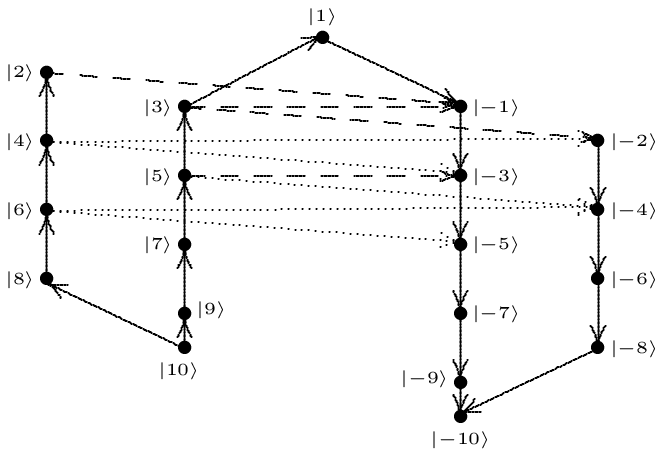}

\end{center}
\caption{Spin relaxation paths (from $m=10$ to $m=-10$)
for $\frac{1}{g\mu_B}\left(A+13B\right)\le H_z\le \frac{2}{g\mu_B}\left(A+34B\right)$. Full lines: thermal transitions due to phonons. Dashed lines: dominant tunneling transitions due to $B_4$ and $H_x$ terms. Dotted lines: tunneling transitions that lead to satellite peaks [included in the numerical diagonalization of the master equation (\protect\ref{veceq})].
}
\label{diagram2}
\end{figure}

\begin{figure}[h]
\begin{center}
    \leavevmode
\epsfxsize=5cm
\epsffile{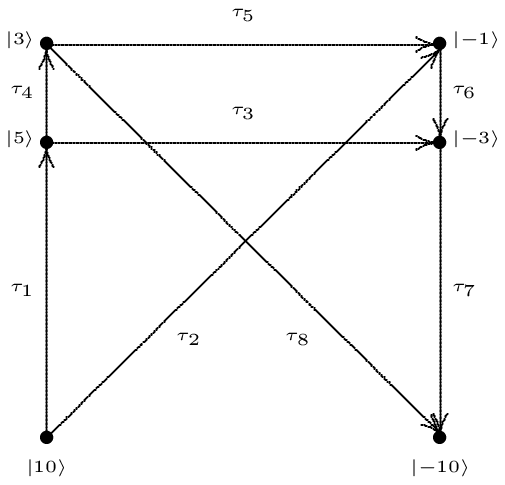}
\end{center}
\caption{Serially reduced diagram associated with Fig.~\protect\ref{diagram2}. In order to understand the analytical evaluation of the relaxation diagram in Fig.~\protect\ref{diagram2} better, tunneling transitions that lead to satellite peaks are excluded.
The relaxation times
$\tau_n$ are given in Eq.~(\protect\ref{analytic}).}
\label{diag2ser}
\end{figure}

$(2)\;\frac{2}{g\mu_B}\left(A+34B\right)\le H_z\le\frac{3}{g\mu_B}\left(A+17B\right)$:
\\
From
$J=J_1+J_2,
J_1=J_3+J_4+J_7,
J_2+J_4=J_5,
J_3+J_5=J_6,
J_6+J_7=J$, and
$\Delta N=J_1\tau_1+J_3\tau_3+J_6\tau_6,
J_3\tau_3=J_4\tau_4+J_5\tau_5,
J_2\tau_2=J_1\tau_1+J_4\tau_4,
J_7\tau_7=J_4\tau_4+J_5\tau_5+J_6\tau_6$,
one can immediately evaluate the relaxation time $\tau_{2,3}^*(H)=\Delta N/J$ (see
Figs.~\ref{diagram3} and \ref{diag3ser}).

\begin{figure}[h]
  \begin{center}
\leavevmode
\epsfxsize=8cm
\epsffile{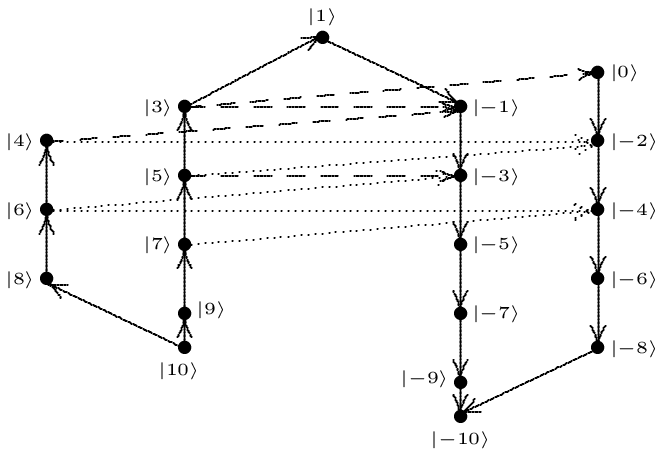}
\end{center}
\caption{Spin relaxation paths (from $m=10$ to $m=-10$)
for $\frac{2}{g\mu_B}\left(A+34B\right)\le H_z\le \frac{3}{g\mu_B}\left(A+17B\right)$.
Full lines: thermal transitions due to phonons. 
Dashed lines: dominant tunneling transitions due to $B_4$ and $H_x$ terms.
Dotted lines: tunneling transitions that lead to satellite peaks [included in the numerical diagonalization of the master equation (\protect\ref{veceq})].}
\label{diagram3}
\end{figure}

\begin{figure}[h]
\begin{center}
    \leavevmode
\epsfxsize=5cm
\epsffile{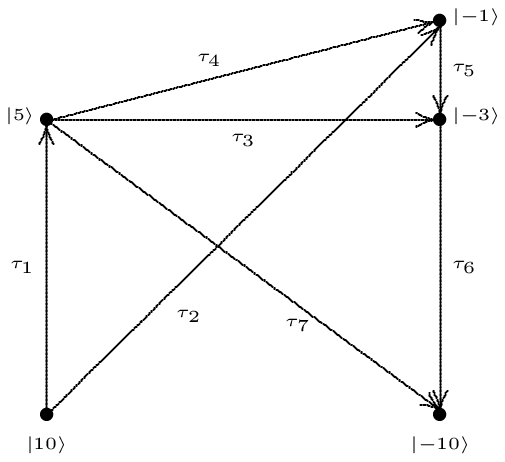}
\end{center}
\caption{Serially reduced diagram associated with Fig.~\protect\ref{diagram3}. In order to understand the analytical evaluation of the relaxation diagram in Fig.~\protect\ref{diagram3} better, tunneling transitions that lead to satellite peaks are excluded. 
The relaxation times $\tau_n$ are given in Eq.~(\protect\ref{analytic}).}
\label{diag3ser}
\end{figure}

$(3)\;\frac{3}{g\mu_B}\left(A+17B\right)\le H_z\le\frac{4}{g\mu_B}\left(A+40B\right)$:
\\
From
$J=J_1+J_2,
J_1=J_3+J_4,
J_4=J_5+J_7,
J_2+J_3+J_5=J_6,
J_6+J_7=J$, and
$\Delta N=J_1\tau_1+J_3\tau_3+J_6\tau_6,
J_3\tau_3=J_4\tau_4+J_5\tau_5,
J_2\tau_2=J_1\tau_1+J_4\tau_4+J_5\tau_5,
J_7\tau_7=J_5\tau_5+J_6\tau_6$,
one can immediately evaluate the relaxation time $\tau_{3,4}^*(H)=\Delta N/J$ (see
Figs.~\ref{diagram4} and \ref{diag4ser}).

\begin{figure}[h]
  \begin{center}
\leavevmode
\epsfxsize=8cm
\epsffile{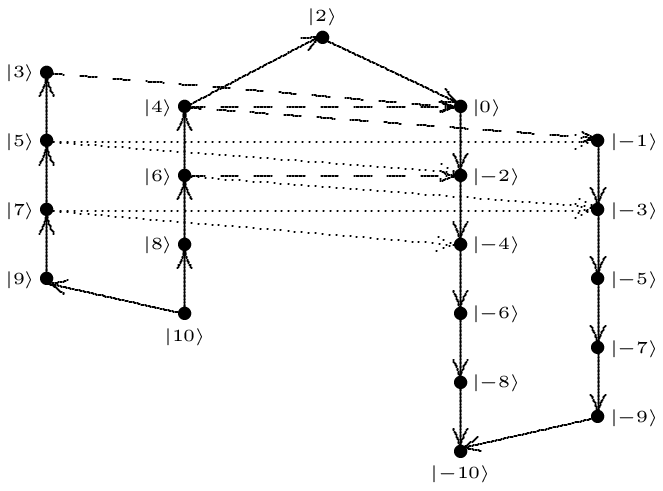}
\end{center}
\caption{Spin relaxation paths (from $m=10$ to $m=-10$) for $\frac{3}{g\mu_B}\left(A+17B\right)\le H_z\le\frac{4}{g\mu_B}\left(A+40B\right)$. 
Full lines: thermal transitions due to phonons. 
Dashed lines: dominant tunneling transitions due to $B_4$ and $H_x$ terms.
Dotted lines: tunneling transitions that lead to satellite peaks [included in the numerical diagonalization of the master equation (\protect\ref{veceq})].
}
\label{diagram4}
\end{figure}

\begin{figure}[h]
\begin{center}
    \leavevmode
\epsfxsize=5cm
\epsffile{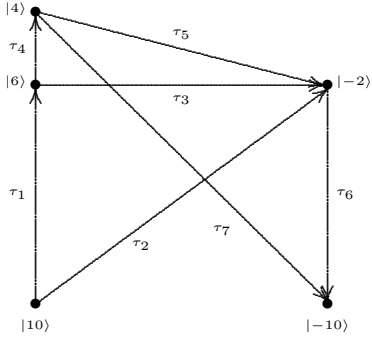}
\end{center}
\caption{Serially reduced diagram associated with Fig.~\protect\ref{diagram4}. In order to understand the analytical evaluation of the relaxation diagram in Fig.~\protect\ref{diagram4} better, tunneling transitions that lead to satellite peaks are excluded. 
The relaxation times $\tau_n$ are given in Eq.~(\protect\ref{analytic}).}
\label{diag4ser}
\end{figure}


\section{First-order vs. second-order transition}
\label{comparison}

We show in this section that second-order transitions lead to a much faster relaxation of the spin system than first-order transitions if the coupling constants are equal. The relaxation rate $\Gamma^{(1)}$ of the cascade with transitions $\Delta m=\pm 1$ has been calculated by Villain {\it et al.} \cite{Villain} (see Fig.~\ref{cascade-fig}),
\bea
\Gamma^{(1)} & = & \frac{3}{2\pi}\frac{|V_{1,0}|^2}{\hbar^4\rho c^5}\left(
\varepsilon_0-\varepsilon_1\right)^3\frac{e^{-\beta\Delta}}{1-e^{-\beta(\varepsilon_0-\varepsilon_1)}} \nn\\
& = &
\frac{3}{2\pi}\frac{|V_{1,0}|^2}{\hbar^4\rho c^5}\left[
\frac{\Delta}{s^2}\right]^3\frac{e^{-\beta\Delta}}{1-e^{-\beta\Delta/s^2}}.
\eea
$\Delta=100A$ is the energy barrier.

\begin{figure}[htb]
\begin{center}
    \leavevmode
\epsfxsize=8cm
\epsffile{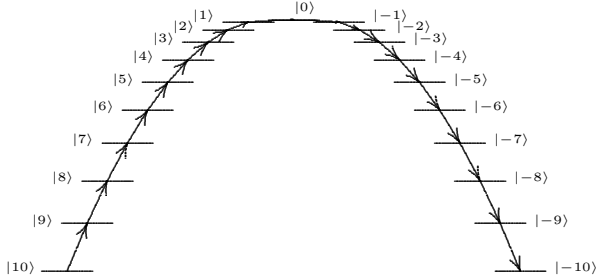}
\end{center}
\caption{Cascade with $\Delta m=-1$ and ${\bf H}=0$.}
\label{cascade-fig}
\end{figure}

We have extended this expression by taking higher-order transitions into account. If we take a cascade with transitions $\Delta m=\pm 2$, for the case s=10, we obtain
\bea
\Gamma^{(2)} & = & \frac{3}{2\pi}\frac{|V_{2,0}|^2}{\hbar^4\rho c^5}\left(
\varepsilon_0-\varepsilon_2\right)^3\frac{e^{-\beta\Delta}}{1-e^{-\beta(\varepsilon_0-\varepsilon_2)}} \nn\\
& = &
\frac{3}{2\pi}\frac{|V_{2,0}|^2}{\hbar^4\rho c^5}\left[
\frac{\Delta}{(s/2)^2}\right]^3\frac{e^{-\beta\Delta}}{1-e^{-\beta\frac{\Delta}{(s/2)^2}}}.
\label{cascade2}
\eea
Comparing to the relaxation rate $\Gamma^{(1)}$ with $s=10$, an increase by a factor
\be
\frac{\Gamma^{(2)}}{\Gamma^{(1)}}\approx\frac{10^6}{5^6}=64
\label{factor}
\ee
is obtained, assuming $V^{(1)}\approx V^{(2)}$ (see Abragam and Bleaney,\cite{Abragam} p.~563, for experimental evidence).

Now we calculate the relaxation rate by means of formula (\ref{relaxation}) with $\Gamma^{(1)}=1/\tau^*$, $\Delta m=\pm 1$, and $\Gamma^{(2)}=1/\tau^*$, $\Delta m=\pm 2$. If there is a fast transition via tunneling between levels $m=4$ and $m'=-4$ for $H_z=0$ at $T=1.9$ K, we get the following more accurate estimation: 
\be 
\frac{\Gamma^{(2)}}{\Gamma^{(1)}}=11.7.
\label{ratio1}
\ee 
The same can be done if the fastest transition takes place via tunneling between levels $m=2$ and $m'=-2$ for $H_z=0$ at $T=1.9$ K,
\be
\frac{\Gamma^{(2)}}{\Gamma^{(1)}}=49.0.
\label{ratio2}
\ee
From these results it is obvious that second-order transitions lead to a faster relaxation. Note that it is Eq.~(\ref{g}) together with Eq.~(\ref{sp_rates}) which imply that the ratios (\ref{ratio1}) and (\ref{ratio2}) are of the same order as the ratio (\ref{factor}). This provides a theoretical justification for the approximation $V^{(1)}\approx V^{(2)}$.

\end{appendix}



\begin{references}
\bibitem{propagator} The free propagator within the integral kernel leads to the $\delta$ distribution in $W_{mn}$ [see Eq.~(\ref{FGR})]. The spin-phonon transition rates (\ref{sp_rates}) are barely affected by the tunnel splitting $E_{mm'}$ [see Eq.~(\ref{E})], which is smaller than the level spacing $\left|\varepsilon_m-\varepsilon_{m'}\right|\ge 0.5$ K (see Sec.~\ref{spin}). Therefore the tunneling part $\cH_{\rm T}$ within the integral kernel can be safely omitted.
\bibitem{Blum_analogy} These expressions are defined in analogy to Eqs.~(8.1.24) and (8.1.29) in Ref.~\onlinecite{Blum}.
\bibitem{detailed_balance} The ratio $W_{m,m+2}/W_{m'-2,m'}$ is approximated by 1. One can see from Eq.~(\ref{sp_rates}) that this approximation is well satisfied near a resonance (i.e., within $w'$).
\end{references}
\end{document}